\documentclass[12pt]{article}
\usepackage{amsfonts}
\usepackage{amssymb}
\usepackage{cite}
\def\hybrid{\topmargin -20pt    \oddsidemargin 0pt
        \headheight 0pt \headsep 0pt
        \textwidth 6.25in       % A4 paper
        \textheight 9.5in       % A4 paper
        \marginparwidth .875in
        \parskip 5pt plus 1pt   \jot = 1.5ex}

\hybrid
\newcommand{\A}{{A_2}}
\newcommand{\B}{{A_1}}
\newcommand{\C}{{A_0}}

\newcommand{\cF}{{\cal F}}

\newcommand{\cM}{{\cal M}}
\newcommand{\cN}{{\cal N}}

\newcommand{\beq}{\begin{equation}}
\newcommand{\eeq}{\end{equation}}
\newcommand{\bi}{\begin{itemize}}
\newcommand{\ei}{\end{itemize}}
\newcommand{\bea}{\begin{eqnarray}}
\newcommand{\eea}{\end{eqnarray}}
\newcommand{\ba}{\begin{array}}
\newcommand{\ea}{\end{array}}
\newcommand{\bt}{\begin{tabular}}
\newcommand{\et}{\end{tabular}}
\newcommand{\bc}{\begin{center}}
\newcommand{\ec}{\end{center}}
\newcommand{\SLtwo}{SL(2,{\bf Z})}
\newcommand{\half}{\frac12}

\def\der{\partial}

\newcommand{\mscr}[1]{\mbox{\scriptsize #1}}

\newcommand{\ft}[2]{{\textstyle\frac{#1}{#2}}}

\def\R{{\rm Re}}
\def\I{{\rm Im}}
\def\theequation{\arabic{section}.\arabic{equation}}

\begin{document}

\begin{titlepage}
\begin{center}

\hfill hep-th/0301125\\
\hfill FSU-TPI-08/02\\
\hfill CERN-TH/2002-377
\vskip 1cm {\large \bf  Effective Actions near
Singularities}\footnote{Work supported by: DFG -- The German
Science Foundation, GIF -- the German--Israeli Foundation for
Scientific Research, European RTN Program HPRN-CT-2000-00148 and
the DAAD -- the German Academic Exchange Service.}

\vskip .5in

{\bf Jan Louis}  \\

{\em II. Institut f\"ur Theoretische Physik,
Universit\"at Hamburg,\\ 
Luruper Chaussee 149, D-22761 Hamburg, Germany}\\
{email: {\tt  jan.louis@desy.de}} \\

\vskip 0.8cm

{\bf Thomas Mohaupt}\\

{\em Theoretisch-Physikalisches Institut,
Friedrich-Schiller-Universit\"{a}t
Jena, Max-Wien-Platz 1, D-07743 Jena, Germany}\\
{email: {\tt  mohaupt@tpi.uni-jena.de}} \\
\vskip 0.5cm

 and \\
\vskip 0.5cm

{\bf Marco Zagermann}\\

{\em CERN, Theory Division,\\ CH-1211 Geneva 23, Switzerland }\\
{email: {\tt  Marco.Zagermann@cern.ch}} \\

\end{center}

\vskip 1.5cm

\begin{center} {\bf ABSTRACT } \end{center}
%\vspace{-2mm}

\noindent 
We study the heterotic string compactified on $K3 \times T^2$
near the line $T=U$, where the effective action becomes
singular due to an $SU(2)$ gauge symmetry enhancement.  
By `integrating in' the light
$W^\pm$ vector multiplets we derive a quantum corrected effective
action which is manifestly $SU(2)$ invariant and non-singular.
This effective action is  
found to be consistent with a residual $SL(2,\mathbf{Z})$
quantum symmetry  on the line $T=U$.
In an appropriate decompactification limit, we recover the known $SU(2)$ 
invariant action in five dimensions.\\

\vfill January 2003

\end{titlepage}

%%%%%%%%%%%%%%%%%%%%%%%%%%%%%%%%%%%%%%%%%%%%%%%%%%%%%%%%%%%%
\setcounter{equation}{0}
%%%%%%%%%%%%%%%%%%%%%%%%%%%%%%%%%%%%%%%%%%%%%%%%%%%%%%%%%%%
%
\section{Introduction}
%
%%%%%%%%%%%%%%%%%%%%%%%%%%%%%%%%%%%%%%%%%%%%%%%%%%%%%%%%%%%%
\setcounter{equation}{0}

The low energy effective action of string theories is of
particular importance, since it captures the interactions of the
light string excitations. Its derivation has a long history and
has been continuously refined. The main idea is to integrate out
the heavy string modes and to derive an effective action of only the
light modes below the mass scale of the heavy excitations.

Generically, this low energy effective action features a set of
moduli scalar fields whose vacuum expectation values are not
determined since they correspond to flat directions of the
effective potential. In compactifications of the ten-dimensional
string theories on compact Ricci-flat manifolds, $Y$, some of the
moduli have a geometrical meaning in that they correspond to
deformations of the metric on $Y$ that preserve the
Ricci-flatness. These deformations can be viewed as coordinates of
the moduli space, $\cM$, of $Y$. Unfortunately, all couplings in
the low energy effective theory depend on these undetermined
vacuum expectation values and therefore phenomenological
predictions are difficult to extract. One expects that
non-perturbative effects generate a potential for the moduli
fields and thus dynamically lift this vacuum degeneracy.

It has become clear for some time that interesting physics is
`hidden' at special points in the moduli space where some
couplings in the effective action become  singular. From a
mathematical point of view, these singularities often arise at
points  (or subspaces) of   the moduli space where the
compactification manifold $Y$ develops a singularity.
 From a physical point of view,
the singularities generically are due to heavy
 fields that become massless at the locus of the singularity.
Integrating these fields out of the effective theory is thus not
legitimate in this region of the moduli space, and this
inconsistency manifests itself as a singularity in some of the
effective couplings.

In general, a consistent, i.e.\ non-singular, effective action
cannot be derived over the entire moduli space, since, as
described above, some of the fields are only light at particular
points (or subspaces) of  the moduli space. Their mass, $M$, is a
nontrivial function of the moduli, and thus $M$ varies over the
moduli space. However, locally near a given singularity  where
some of the fields are light and $M$ approaches zero, one can
choose to not integrate out these light fields and derive a
non-singular effective action in the vicinity of  $M=0$, i.e.\
near the region of the former singularity. Differently put -- and
that is the way we will proceed in this paper -- one can start
from the singular action and locally `integrate the light modes
back in'.\footnote{The term `integrating in' was coined in
\cite{Intriligator} in a slightly different context.}

The purpose of this paper is to perform this `integrating in'
procedure in some detail in a model where the
 singular
effective action is known exactly. More specifically, we consider
the heterotic string compactified on $K3\times T^2$, which leads
to an effective theory with $N=2$ supersymmetry in four space-time
dimensions ($d=4$). This class of string backgrounds is believed
to be dual to type IIA string theory compactified on $K3$ fibered
Calabi-Yau threefolds 
\cite{KV,FHSV,KLM,AL,CCLM,LF}. As a consequence,
some of the couplings of the low energy effective theory are known
exactly.

The low energy $N=2$ supergravity theory  contains, apart from the
gravitational multiplet, a set of $n_V$ vector multiplets and
$n_H$ hypermultiplets. Both multiplets contain scalar fields which
can be viewed as the coordinates of the moduli space $\cM$. As a
consequence of $N=2$ supersymmetry, this moduli space factorizes:
$\cM = \cM_V\times \cM_H$, where $\cM_V$ is spanned by the scalars
in the vector multiplets, and $\cM_H$ is spanned by the scalars in
the hypermultiplets. Due to this factorization, one can discuss
each component separately, and in this paper we will only focus on
$\cM_V$. $\cM_V$ is constrained to be a special K\"ahler manifold
\cite{wp,N=2}, that is, $\cM_V$ is endowed with a K\"ahler metric
which  can be expressed in terms of a holomorphic prepotential
$\cF$ (see Appendix A).

For a certain class of compactifications, $\cF$ is known exactly,
and here we focus on a very specific model known as the
$STU$-model. It corresponds to a compactification of the heterotic
string on a $K3 \times T^2$ manifold with instanton numbers
$(14,10)$. This model   is non-perturbatively dual to the IIA
string compactified on the Calabi-Yau threefold
$Y_{1,1,2,8,12}(24)$, which is an elliptic fibration over the
second Hirzebruch surface ${\bf I\!F}_2$ 
\cite{MorVaf1,Wit:96,LSTY}. 
The generic low energy
effective theory contains $n_V=3$ vector multiplets (whose complex
scalar fields we denote by
  $S,T,U$)  and $n_H=244$ neutral hypermultiplets. The hypermultiplets
 play no role
in the following and will be consistently ignored. The gauge group
at a generic point in the moduli space is $G=U(1)^4$, where the
additional gauge boson is the $N=2$ graviphoton. At the subspace
$T=U$, $G$ is enlarged to $G=U(1)^3 \times SU(2)$, at $T=U =1$ it
is further enhanced to $G=U(1)^2 \times SU(2)^2$, and at $T=U
=\rho\equiv e^{i\pi/6}$  one finally has $G=U(1)^2 \times SU(3)$.
In other words, at these special points additional gauge bosons
(or rather $N=2$ vector multiplets) become massless and enhance
the everywhere existing Abelian to a non-Abelian gauge symmetry.
From  the point of view of the heterotic string, this is the usual
perturbative gauge symmetry enhancement due to additional massless
Kaluza-Klein and winding modes for particular values of
the moduli of the two-torus. 
This symmetry enhancement does not survive
non-perturbative quantum corrections \cite{SW,KLM}, and we will
therefore restrict our considerations to the perturbative
heterotic string only.\footnote{From the point of view of the type
IIA string, the $SU(2)$ gauge symmetry enhancement is, at the
classical level, due to a shrunken two-cycle of the Calabi Yau
three-fold,
  with the wrapped D2 brane
giving rise to the $W^{\pm}$ bosons. Unlike its M-theory analogue
\cite{AntFerTay,Wit:96,MZ},
this geometrical singularity (and hence the symmetry enhancement)
does not survive quantum corrections. These quantum corrections
can be calculated in the dual IIB picture using mirror symmetry. 
The mirror threefold develops a conifold singularity and one 
gets massless hypermultiplets with magnetic or dyonic charge,
which come from D3 branes wrapping the vanishing three-cycle \cite{Str:95}.}

The one-loop prepotential $\cF^{(1)}$ is singular at $T=U$, which
signals the existence of the additional light states. As we show
in this paper, it is possible to derive an effective action valid
near $T=U$ which is non-singular and which contains the $W^\pm$
gauge bosons of the $SU(2)$.  Following the approach of \cite{MZ},
we   will not do this via a microscopic string theory calculation,
but rather
  by using symmetry arguments to
reconstruct the non-singular theory from the well-known
\cite{DKLL,AFGNT,HM}
 singular effective action, in which the $W^{\pm}$ bosons are integrated out.

Our motivation for this work is to study compactifications of 
string theory and  
M-theory in situations where the effective action becomes
singular due to the presence of additional light modes. Such 
extra states might be either of perturbative or non-perturbative
origin. In this paper, we consider the heterotic perturbative mechanism
of $SU(2)$ enhancement, as explained above. The natural next 
step will be to consider conifold singularities in compactifications
of type II string theory on Calabi-Yau threefolds \cite{Str:95}. In this case
the additional states are non-perturbative and descend from 
D-branes wrapped on a vanishing cycle. Since heterotic string compactifications 
on $K3\times T^2$ are dual to type II compactifications 
on Calabi-Yau threefolds,
we expect that the results of this
paper will be useful for the study of conifolds.

There are further, even more interesting cases one might wish to
consider. One line of developement will be the study of extremal
transitions, where one has, in contrast to conifold transitions,
an unbroken non-abelian gauge symmetry at the transition point
\cite{KatMorPle,KleEtAl}. Another interesting extension is to add
background flux, which can resolve the singularity
\cite{KleStr:00} and create a hierarchically small scale
\cite{Mayr,GidKacPol:01}. Besides Calabi-Yau compactifications one
should also try to study  $N=1$ supersymmetric
 $G_2$-compactifications of M-theory along similar lines.
Here the understanding of singular manifolds is mandatory,
  as smooth $G_2$-compactification
do not lead to non-abelian gauge groups or chiral fermions
\cite{PapTow:95}. We hope to return to these issues in later
publications.

This paper is organized as follows. In section~\ref{STUreview}, we
briefly recall the results for the perturbative prepotential in
the $STU$-model. In section~\ref{Stree}, we then derive the tree
level action for the effective theory near $T=U$ with the $W^\pm$
gauge bosons (and their superpartners) included. We explicitly
compute the potential and the masses of the $N=2$ $W^\pm$
supermultiplets. In section~\ref{Sloop} we determine the one-loop
corrections to this effective action near $T=U$. Based on general
arguments, we first show (section \ref{Sloopgen}) that the minimum
of the quantum corrected potential does not change and the masses
only receive multiplicative corrections which are entirely due to
corrections of the K\"ahler potential. In analogy with $N=1$
supergravity, it is possible to define a  `holomorphic mass' which
remains uncorrected. In section~\ref{Floop}, we determine the
loop-corrected prepotential by using the known result of the
$STU$-model. We `undo' the integrating out procedure of the
$W^\pm$ gauge bosons by subtracting their threshold corrections to
the $SU(2)$ gauge couplings. This way we derive a  non-singular
quantum corrected prepotential for the $SU(2)$ gauge theory. In
section~\ref{qsymm} we check that this prepotential transforms
appropriately under the expected residual quantum duality symmetry
$SL(2,{\bf Z})$ and that it does have the proper singularities at
points in the moduli space where further gauge enhancement occurs.
This is an independent check on our procedure. Finally, in
section~\ref{largeR}, we decompactify the theory to five
space-time dimensions and establish the consistency with the
results of \cite{MZ}. 
Some technical details are relegated to three appendices. In
appendix \ref{sugra}, we supply the necessary facts of $N=2$
supergravity. In appendix \ref{polylog}, we assemble some useful
formulae about the polylog series, while in appendix \ref{smodform}
we review properties of modular forms.

%%%%%%%%%%%%%%%%%%%%%%%%%%%%%%%%%%%%%%%%%%%%%%%%%%%%%%%%%%%
\section{Preliminaries: Review of the $STU$-model}\label{STUreview}
\setcounter{equation}{0}

Let us first recall a few facts about the $STU$-model
\cite{DKLL,AFGNT,HM,KLM,LF,CarLueMoh:95,CCLM}. At the string tree level it
is characterized by the prepotential\footnote{For a review of
$N=2$ supergravity see appendix~A.} \beq\label{Fst} \cF^{(0)} = -
STU = - S(T_+^2 -T_-^2)\ , \eeq where $T_\pm \equiv \frac1{2}(T\pm
U)$.
%
% Inserted into (\ref{Kspecial}) one obtains the K\"ahler potential
% \bea
% K^{(0)}&=& -\log\Big[(S+\bar S)(T+\bar T)(U+\bar U)\Big] \\
% &=& -\log(S+\bar S)-\log\Big[(T_++\bar T_+)^2 - (T_-+\bar T_-)^2\Big]
% \ .\nonumber
% \eea
%
The quantum correction  of the vector multiplet couplings can be
parameterized by corrections to this prepotential. They only
appear at 1-loop (generating a correction  $\cF^{(1)}$) and
non-perturbatively (generating a contribution $\cF^{(NP)}$).
Therefore, the quantum corrected  $N=2$ prepotential obeys the
expansion $\cF = \cF^{(0)}+ \cF^{(1)}+ \cF^{(NP)}$. $\cF^{(1)}$ is
known from a heterotic computation \cite{DKLL,AFGNT,HM}, while
$\cF^{(NP)}$ is known from the duality to IIA on the Calabi-Yau
threefold
 $Y_{1,1,2,8,12}(24)$ \cite{KLM,HM}.
As we restrict ourselves to the  perturbative heterotic string,
only the one-loop correction $\cF^{(1)}$ is of interest. For the
rest of this paper we neglect the non-perturbative correction
$\mathcal{F}^{(NP)}$ and, by abuse of notation, simply write
\begin{equation}
\cF=\cF^{(0)}+\cF^{(1)} \ .
\end{equation}

We already displayed $\cF^{(0)}$ in (\ref{Fst}). For
$\textrm{Re }T>\ \textrm{Re }U$, $\cF^{(1)}$ is  given by \cite{HM}
\begin{equation}
\cF^{(1)} =  -\ft1{12\pi} U^3 - \ft{1}{(2 \pi)^4} Li_3 \left(
e^{-2 \pi (T-U)} \right)  - \ft{1}{(2 \pi)^4}
\sum_{k,l=0}^{\infty} c_1(kl) Li_3 \left( e^{-2\pi (kT + lU)}
\right)\ , \label{WeylChamber1}
\end{equation}
where the third polylog $Li_3$ is defined in appendix~B, and the
coefficients
  $c_1(kl)$
can be found in \cite{HM}. As was pointed out in
\cite{CFG,DKLL,HM}, $\cF^{(1)}$ is only determined up to a
quadratic polynomial in the variables $1,iT,iU,TU$ with purely
imaginary coefficients. Adding any such polynomial simply amounts
to a shift in  theta angles; we will come back to this ambiguity
in section~\ref{Floop}.

$\cF^{(1)}$ is largely determined by its quantum symmetries. The
$STU$-model has the perturbative quantum symmetry
$SO(2,2;\mathbf{Z})$, which
 includes the exchange
$\sigma: T\leftrightarrow U$ as well as the duality  group
$\SLtwo_{T}\times\SLtwo_{U}$. Here, $\SLtwo_{T}$ acts on $T,U$  as
\beq T\to {aT-ib\over icT+d}\ ,\qquad U\to U\ , \qquad ad-bc =1,
\quad a,b,c,d, \in {\bf Z}\ , \eeq whereas the action of
$\SLtwo_{U}$   is obtained by  exchanging $T$ with $U$. As a
consequence of this symmetry, the third derivative $\partial_T^3
\cF^{(1)}$ is a modular form of weight $(+4,-2)$, while
$\partial_U^3 \cF^{(1)}$ is a modular form of weight $(-2,+4)$
under $\SLtwo_{T}\times\SLtwo_{U}$. They are given by
\cite{DKLL,AFGNT} \bea
\partial_T^3\cF^{(1)} &
=& {+1\over2\pi}\,{E_4(iT)\, E_4(iU) E_6(iU) \eta^{-24}(iU)
        \over j(iT)\,-\,j(iU)}\,,\nonumber\\
\partial_U^3\cF^{(1)} &
=& {-1\over2\pi}\,{E_4(iU) E_4(iT) E_6(iT) \eta^{-24}(iT)
        \over j(iT)\,-\,j(iU)}\,,
\label{hfinal1} \eea where  the modular forms $E_4$, $E_6$,
$\eta$, $j$ are defined in Appendix C.

As we discuss more explicitly in sections~\ref{Sloop} and
\ref{qsymm},
  $\cF^{(1)}$
 is singular at $T=U$ due to gauge symmetry enhancement:
\bea
T=U:\quad U(1)\times U(1) &\to& U(1)\times SU(2)\ ,\nonumber\\
T=U=1:\quad U(1)\times U(1) &\to& SU(2)\times SU(2)\ ,\\
T=U=\rho:\quad U(1)\times U(1) &\to& SU(3)\ .\nonumber \eea At
these points additional massless vector multiplets appear which
should not have   been integrated out of the effective action and
which are the origin of the singular couplings (\ref{hfinal1}).

%%%%%%%%%%%%%%%%%%%%%%%%%%%%%%%%%%%%%%%%%%%%%%%%%%%%%%%%%%%%%%%%%%%%

%%%%%%%%%%%%%%%%%%%%%%%%%%%%%%%%%%%%%%%%%%%%%%%%%%%%%%%%%%%%%%%%%%%%

\section{The tree level effective action }
\label{Stree} \setcounter{equation}{0} Our goal in this paper is
to derive a non-singular effective action that gives an accurate
description of the theory near the surface $T=U$, where the
$SU(2)$ gauge symmetry enhancement occurs. In this section, we
restrict ourselves to the tree level approximation of this
effective theory. This sets our notation and prepares the
discussion of the one-loop corrections.

%Before we derive the detailed form of this theory in Sections 5
%and 6, we will use this section to collect a few general
%statements that can be made without much effort and without the
%explicit knowledge
% of all the  couplings.
% Most of these statements are physically rather obvious,
%but it is nevertheless instructing to see them embedded in the
%formalism of special K\"{a}hler geometry.

Let us begin with our notation regarding the spectrum. Near the
surface $T=U$, the set of light fields in the low energy effective
 action
has to be enlarged to also include  the $W^{\pm}$ bosons (along
with  their superpartners). The effective action is thus an $N=2$
supergravity theory with $3+2=5$ vector multiplets in which
$SU(2)$ is realized as a Yang-Mills-type gauge symmetry.
Three of these five vector multiplets have to transform in the
adjoint representation of $SU(2)$, and we  use $C^a,  a=1,2,3$, to
denote the complex scalar fields of this triplet.
We choose to identify $C^1$ and $C^2$ with the scalar
superpartners of the $W^{\pm}$ bosons:
\begin{equation}
W^{\pm}=C^1\pm iC^2.
\end{equation}
The scalar field $C^3$ then has to be identified with $T_{-}=\half
(T-U)$, whose vacuum expectation value triggers the symmetry
breaking $SU(2)  \longrightarrow U(1)$    via a supersymmetric
Higgs effect.
In addition to the triplet $C^{a}$, there are two $SU(2)$ singlet
vector multiplets, and  at tree level the scalars of these singlet
multiplets can be chosen to coincide with the moduli $S$ and
$T_{+}$.\footnote{%
In general, the two singlets have to be invariant under the Weyl
twist $C^{3} \rightarrow -C^{3}$, which is the only remnant of the
$SU(2)$ symmetry after $C^{1}$ and $C^{2}$ have been integrated
out. According to the identification $C^{3}=T_{-}$, the Weyl twist
is equivalent the exchange symmetry $\sigma:T \leftrightarrow U$.
At string tree level, both $S$ and $T_{+}$ are $\sigma$-invariant
and can therefore be identified with the two $SU(2)$ singlets, as
we did above.}
% At the 1-loop level, however, $S$ becomes a multivalued
% function on the moduli space and transforms non-trivially under the
% perturbative duality group $SO(2,2;\mathbf{Z})$ \cite{DKLL}. In
% particular, it transforms non-trivially under $\sigma$ and can
% therefore no longer serve as an $SU(2)$ singlet. At the 1-loop
% level,  $S$ therefore has to be replaced by a suitable
% $\sigma$-invariant linear combination, $\hat{S}$, of the moduli
% $S,T_+,T_-$, and the appropriate $SU(2)$ singlet scalars are given
% by $(\hat{S},T_+)$. We will come back to this point in Section~\ref{Sloopgen}
% when we discuss the 1-loop corrections.}
%
The   scalar fields $(C^a,S,T_+)$ are  `special' coordinates of a
symplectic section $(X^I,F_J)$ $(I,J=0,1,\ldots, 5)$, which in our
conventions (see Appendix A for details) means that
\begin{equation}\label{variables}
\frac{X^{j}}{X^{0}}=t^{j}=(iC^a,iS,iT_{+})\ ,  \qquad j=1,\ldots,5\ .
\end{equation}

In the following, we  use the subscript `in' to label all actions,
prepotentials, etc., where the two $W^{\pm}$ bosons have been
`integrated in'. More explicitly, $S_{\rm in}$ denotes the full
perturbative effective action near $T=U$  with the $W^{\pm}$
bosons included, while $\mathcal{F}_{\rm in}$ refers to the
underlying prepotential. The corresponding tree level quantities
%we plan to determine  in this section
are denoted by $S_{\rm in}^{(0)}$ and $\mathcal{F}_{\rm
in}^{(0)}$, respectively.

%The defining property of $S_{\rm in}^{(0)}$ is that it reproduces
%the tree level action $S^{(0)}$ encoded in the prepotential
%$\mathcal{F}^{(0)}$

The prepotential $\mathcal{F}_{\rm in}^{(0)}$ is a holomorphic
function of the variables  $(C^a,S,T_+)$. Its defining property is
that the corresponding action, $S_{\rm in}^{(0)}[C^a,S,T_+]$,
should reproduce the action $S^{(0)}[S,T_+,T_-]$ encoded in the
prepotential $\cF^{(0)} = - S(T_+^2 -T_-^2)$ of eq.\ (\ref{Fst}),
when $C^{1,2}$ (and their superpartners) are integrated out.
At tree level, no threshold effects can occur, and integrating out
these two multiplets simply means to set them equal to zero in the
action $S_{\rm in}^{(0)}[C^a,S,T_+]$ or, equivalently, in the
prepotential $\mathcal{F}_{\rm in}^{(0)}(C^a,S,T_+)$.\footnote{The
consistency of this truncation is guaranteed because $C^{1,2}$
form a doublet of an obvious $SO(2)\subset SU(2)$.} As the
`in-theory' is to be $SU(2)$ invariant, the triplet $C^a$ can only
appear via its $SU(2)$ invariant combination $C^a C^a$, and
integrating out $C^{1,2}$ at tree level is thus tantamount  to
making the replacement
\begin{equation}
(C^a C^a)\longrightarrow (T_-)^2
\end{equation}
everywhere in $\mathcal{F}_{\rm in}^{(0)}(C^a,S,T_{+})$.
Conversely,
  $\mathcal{F}_{\rm in}^{(0)}(C^a,S,T_+)$ can simply be
obtained from $\cF^{(0)}(S,T_+,T_-)   = - S(T_+^2 -T_-^2)$ by  the
inverse substitution
\begin{equation}
(T_-)^2 \longrightarrow (C^a C^a).
\end{equation}
We therefore arrive at
\begin{equation}\label{Ftree}
\mathcal{F}_{\rm in}^{(0)}=-S[T_{+}^2-C^a C^a]
\end{equation}
as the tree level prepotential with the two $W^{\pm}$ bosons
included.

As a consequence of $N=2$ supersymmetry,
% the tree-level effective action 
$S_{\rm in}^{(0)}[C^a,S,T_+]$ is
completely determined by (\ref{Ftree})  and can in principle be
worked out in all detail using the relations  reviewed in Appendix
A. For the rest of this section, let us content ourselves with a
short discussion of some of the quantities that play a role in
the integrating out process.

Consider first   the K\"{a}hler potential, $K$, of the scalar
manifold, $\mathcal{M}_V$. Using (\ref{Kspec}),  one obtains for
$K$
 \beq\label{KSU2} K= -\log (S+\bar S) -\log Y\
,\qquad Y\equiv (T_+ +\bar T_+)^2 - (C^a+\bar C^a) (C^a+\bar C^a)\
. \eeq This is the
  K\"{a}hler potential of  the symmetric space $\mathcal{M}_{V}=
\frac{SU(1,1)}{U(1)}\times \frac{SO(2,4)}{SO(2)\times SO(4)}$ with
isometry group $\textrm{ISO}(\mathcal{M}_{V})=SU(1,1)\times
SO(2,4)$. The form  (\ref{KSU2}) corresponds  to a parametrization
in which only the subgroup $SO(1,1)\times SO(1,3)$ is  a manifest
symmtry of the K\"{a}hler potential.\footnote{The $SO(1,1)$ factor
acts as $S\rightarrow \lambda^2 S$, $(C^a,T_{+})
 \rightarrow (\lambda^{-1}C^{a},\lambda^{-1}T_{+})$.}
The Yang-Mills-type gauge group of the theory is to be identified
with the $SU(2)$ subgroup of the $SO(1,3)$ factor. On the
homogeneous coordinates, $X^{I}$, this $SU(2)$ acts as
\begin{eqnarray}
\delta X^{a}&=&\Lambda^b \epsilon_{bca}X^{c}\nonumber\\
\delta X^{0}&=&\delta X^{4}\, =\, \delta X^{5}\, =\,
0,\label{SU2trafo}
\end{eqnarray}
where we have identified the structure constants
$f_{ab}^c=\epsilon_{abc}$. On the scalar manifold,
$\mathcal{M}_{V}$, the corresponding  $SU(2)$ isometries are
generated by Killing vectors $(k_{b}^{a},k_{b}^{+},k_{b}^{S})$:
\begin{eqnarray}
\delta C^{a}&=&\Lambda^{b}k_{b}^{a}\nonumber\\
\delta T_{+}&=&\Lambda^{b}k_{b}^{+}\nonumber\\
\delta S &=&\Lambda^{b} k_{b}^{S}.\label{Killingtrafo}
\end{eqnarray}
From the relation (\ref{variables}),
  one   reads off\footnote{Using eqs. (\ref{kP}) and (\ref{PF}),
one arrives at  the same result.}
\begin{equation}\label{PKSU2}
k_{a}^{b}=\epsilon_{abc}C^{c}, \qquad k_{a}^{+}=k_{a}^{S}=0 \ .
\end{equation}
The Killing vectors enter the covariant derivatives of the scalar
fields
 (see (\ref{gaugeco})),
%\begin{equation}
% D_\mu C^a =
%\partial_\mu C^a - k^{a}_b A_\mu^b \,
%\end{equation}
as well as the scalar potential,
\begin{equation}\label{Vpot1}
 V = 2 \ e^K (X^I k_I^{\bar\imath})  g_{\bar \imath j}\, (\bar X^J k_J^j) \ .
\end{equation}
Obviously,    $V$ is   positive semi-definite, and zero if and
only if
\begin{equation}\label{condition}
(\bar X^J k_J^j)=0.
\end{equation}
In view of (\ref{PKSU2}), this means that the vacua of the theory
correspond to field configurations with
\begin{equation}
[C,C^{\dagger}]=0, \qquad \textrm{where } C\equiv C^a\sigma^a .
\end{equation}
Thus, any vacuum can be brought to the form $\langle C^1 \rangle =
\langle C^2 \rangle = 0$ by means of an $SU(2)$ transformation. As
the gaugino variations are proportional to the quantity
 $(\bar X^J k_J^j)$ \cite{wp,N=2},
all these vacua also preserve the $N=2$ supersymmetry (and also
exhaust all $N=2$ supersymmetric Minkowski vacua).

Remembering $C^3=T_{-}$, we have therefore, at tree level,
 simply rediscovered
that, modulo $SU(2)$ transformations, $(T_{-},S,T_{+})$ indeed
parametrize the flat directions  of the scalar potential
 and that all of the corresponding vacua are $N=2$ supersymmetric.

Let us close this section with a short discussion of the
tree-level masses of the $W^{\pm}$ bosons and their scalar
superpartners in these vacua. The mass of any scalar field arises
from non-vanishing second derivatives of the scalar potential.
Combining (\ref{KSU2}), (\ref{PKSU2}) and (\ref{Vpot1}), the
tree-level scalar potential is
\begin{eqnarray}
V&=&e^K \frac{4}{Y}[(\bar{C}^a C^a)^2-(\bar{C}^a\bar{C}^a)(C^b C^b)]\nonumber\\
&=&\frac{4}{(S+\bar{S})Y^2}[(\bar{C}^a
C^a)^2-(\bar{C}^a\bar{C}^a)(C^b C^b)]
\nonumber\\
&=&\frac{1}{2(S+\bar{S})Y^2}\textrm{tr} ([C,C^{\dagger}]^2).
\end{eqnarray}
At $\langle C^{1} \rangle =\langle C^{2}  \rangle =0$, the only
non-vanishing  second derivatives of this potential   are
\begin{eqnarray}
\partial_{1}\partial_{1}
V\big|_{ \langle C^1 \rangle =  \langle C^2 \rangle  =0} &=&
- e^{K} \frac{8}{Y} ({\bar{C}^3})^2\nonumber\\
\partial_{1}\partial_{\bar{1}}
V\big|_{ \langle C^1 \rangle =  \langle C^2 \rangle  =0} &=&
 e^{K} \frac{8}{Y} |C^3|^2\nonumber\\
\partial_{\bar{1}}\partial_{\bar{1}}
V\big|_{ \langle C^1 \rangle =  \langle C^2 \rangle  =0} &=& -
e^{K} \frac{8}{Y} (C^3)^2\
\end{eqnarray}
and similarly for the derivatives with respect to $C^2$ and
${\bar{C}}^2$.

In order to diagonalize these mass matrices, one decomposes $C^1$
and $C^2$ \ into the real fields parallel and perpendicular to
$\langle C^3 \rangle$ :
\begin{eqnarray}
C^1(x)&=&a^1(x) \langle C^3 \rangle +i b^1(x) \langle C^3 \rangle\nonumber\\
C^2(x)&=&a^2(x) \langle C^3 \rangle +i b^2(x) \langle C^3 \rangle
. \label{decompo}
\end{eqnarray}
One then finds
\begin{equation}
\frac{\partial^2 V}{\partial b^1 \partial b^1} \Big|_{ \langle C^1
\rangle =  \langle C^2 \rangle  =0}= \frac{\partial^2 V}{\partial
b^2 \partial b^2} \Big|_{ \langle C^1 \rangle =  \langle C^2
\rangle  =0} = 32 \frac{e^K}{Y}|C^3|^4
\end{equation}
with all other derivatives vanishing. Taking into account
\begin{eqnarray}\label{g11g22first}
g_{1\bar{j}}\big|_{ \langle C^1 \rangle =  \langle C^2 \rangle
=0}
&=&\delta_{1 j}\, \frac{2}{Y}\ ,\nonumber\\
 g_{2\bar{j}}\big|_{ \langle C^1 \rangle =  \langle C^2 \rangle  =0}
&=&\delta_{2 j}\, \frac{2}{Y}\ ,
\end{eqnarray}
the kinetic terms of $C^{1}$ and $C^{2}$ simplify to
\begin{eqnarray}
\mathcal{L}_{\rm kin}&=& -\frac{2}{Y}\, \partial_{\mu}C^1
\partial^{\mu} \bar{C^1}
-\frac{2}{Y}\, \partial_{\mu}C^2 \partial^{\mu} \bar{C^2}\nonumber\\
&=& -\frac{2}{Y}\, |C^3|^2 (\partial_{\mu}a^1 \partial^{\mu} a^1
+\partial_{\mu}b^1 \partial^{\mu} b^1 +\partial_{\mu}a^2
\partial^{\mu} a^2 +\partial_{\mu}b^2 \partial^{\mu} b^2),
 \end{eqnarray}
and we see that only $b^{1,2}$ obtain a mass, but  not $a^{1,2}$
or any other scalar field. The mass of the corresponding
canonically normalized scalar fields is given by
\begin{equation}\label{scalarmass}
M^2=8 e^{K} |C^3|^2=8 e^{K}|T_{-}|^{2}.
\end{equation}
This mass formula agrees with the one obtained from string theory,
which reads \cite{CLM}:
\beq
M^2 = \frac{4}{\alpha'} \frac{ |T-U|^2 }{(T+ \overline{T})
(U+\overline{U})} = 
\frac{16}{\alpha'} \frac{|T_-|^2}{Y} \;.
\eeq
To see the agreement, one has to reinstall the gravitational coupling
$\kappa$, which we have set to unity throughout, to
convert string units into gravitational
units using \cite{Kap:92} $g^2 \alpha' \kappa = 4$ and to 
express the heterotic string coupling  $g$ 
through the vev of the dilaton, $\ft1{g^2} = \ft{\langle S +
\overline{S} \rangle}{2}$. 

In analogy with $N=1$ theories, we define a holomorphic mass, $m$,
through the relation $M^2 =8 e^{K} |m(T)|^2$,
 which for the case at hand implies\footnote{%
The mass parameters of an $N=1$ superpotential are necessarily
holomorphic and a similar feature holds for masses generated via a
supersymmetric Higgs effect \cite{KL}. Such holomorphic mass
parameters are of importance due to their non-renormalization
properties. In $N=2$ theories one can analogously define a
holomorphic mass, and, as we will see in section~\ref{Sloopgen},
this mass is not renormalized.} 
\begin{equation}\label{holomorphicmscalar}
m= T_{-}\ .
\end{equation}

As the
  vacua with $\langle C^1 \rangle = \langle C^2 \rangle  =0$ preserve the full
$N=2$ supersymmetry, one should observe     a supersymmetric Higgs
effect in which the vector fields $A_{\mu}^{1,2}$ (i.e., the
$W^{\pm}$ bosons) absorb the massless components $a^{1,2}$ and
acquire the same mass as their scalar superpartners $b^{1,2}$
(\ref{scalarmass}). And indeed, the mass term for the vector
fields arises from the square of the covariant derivative \beq
D_\mu C^a =
\partial_\mu C^a - k^{a}_b A_\mu^b \, \eeq which leads to the mass
matrix \beq\label{MSU2first} M^2_{ab} = 2 g^2 g_{c\bar d} k^c_a
\bar k^{\bar d}_b \, \eeq where $g^2=-2\langle \textrm{Im}
\mathcal{N}_{11}\rangle^{ -1}=-2\langle \textrm{Im}
\mathcal{N}_{22}\rangle^{-1}$ ensures the correct canonical
normalization. For  $\langle C^1 \rangle =  \langle C^2 \rangle
=0$, one has 
\begin{equation}
g^{-2}=-\frac{ \langle\textrm{Im} \mathcal{N}_{11} \rangle}{2}=
-\frac{\langle \textrm{Im}
\mathcal{N}_{22} \rangle}{2} = \frac{\langle S+\bar{S} \rangle}{2}
\end{equation}
so that one  indeed   obtains the same mass matrices as above:
\begin{eqnarray}
M^2_{ab}  &=&8 e^K\ {\rm diag} (|C^3|^2,|C^3|^2,0)
= 8 e^K\ |T_-|^2\ {\rm diag} (1,1,0) \label{MWfirst}\\
m_{ab}&=&   T_- \ {\rm diag} (1,1,0). \label{holomorphi}
\end{eqnarray}
As $m$ depends holomorphically on the moduli, it
should not receive  loop corrections. We will verify this in
Section~\ref{Sloopgen}.

%%%%%%%%%%%%%%%%%%%%%%%%%%%%%%%%%%%%%%%%%%%%%%%%%%%%%%%%%%%%%%%%%%

%%%%%%%%%%%%%%%%%%%%%%%%%%%%%%%%%%%%%%%%%%%%%%%%%%%%%%%%%%%%%%%%%%%

\section{The one-loop corrections}\label{Sloop}
\setcounter{equation}{0} In this section we go beyond the tree
level approximation and determine the one-loop corrections to the
effective action $S_{\rm in}^{(0)}$ described in the previous
section. Our result will be an action $S_{\rm in}$ that describes
the full low energy dynamics of the perturbative heterotic string
near $T=U$. This action again involves the coupling of five vector
multiplets to $N=2$ supergravity and exhibits an $SU(2)$ gauge
symmetry. Several properties of this action follow already from
its gauge invariance and can  be inferred without a detailed
knowledge of all the couplings. We list these general properties
in section~\ref{Sloopgen} before we construct the complete theory
with all the detailed couplings in section~\ref{Floop}.

\subsection{Some general properties of $S_{\rm in}$}\label{Sloopgen}
Just as in the tree level case, three of the five vector fields
transform in the adjoint representation of $SU(2)$, while the
remaining two have to be $SU(2)$ inert. The scalars of the triplet
are again denoted by $C^{a}$ $(a=1,2,3)$, with $C^{1,2}$
corresponding to the $W^{\pm}$ bosons and $C^{3}=T_{-}$. In the
tree level approximation, we could choose $S$ and $T_{+}$ as the
scalar fields of the two singlet multiplets, because both are
classically invariant under the Weyl twist $\sigma: T
\leftrightarrow U $, which is the only remnant of the $SU(2)$
gauge symmetry once the $C^{1,2}$ are  integrated out.  At one
loop, this is still true for $T_{+}$, however,
 $S$ now  becomes a multivalued
function on the moduli space and transforms non-trivially under
the perturbative duality group $SO(2,2;\mathbf{Z})$ \cite{DKLL}.
More precisely, using eq.\ (4.27) in ref. \cite{DKLL}, one finds
\begin{equation}\label{Strafo}
\sigma: S\longrightarrow S-\frac{1}{2\pi} T_{-}\ .
\end{equation}
Thus, $S$ can no longer serve as one of the $SU(2)$ singlets.
Fortunately, it is easy to construct a $\sigma$-invariant linear
combination  out of $\{S,T_{+},T_{-}\}$:\footnote{The most general
$\sigma$-invariant linear combination would be $\hat{S}+aT_{+}$
with $a$ arbitrary. $\hat{S}$ is singled out   by the property
$\hat{S}\big|_{T_{-}=0}=S\big|_{T_{-}=0}$, a property
    that simplifies some of the equations in Section \ref{qsymm}.
Note that neither $\hat{S}$ nor $\hat{S}+aT_{+}$ is the `invariant
dilaton' $S^{inv}$ described in \cite{DKLL}. The invariant dilaton
$S^{inv}$
  is a highly non-linear
  function of the moduli that is invariant under the full duality group
$SO(2,2;\mathbf{Z})$. $\hat{S}$, by contrast,  is only invariant
under the Weyl twist $\sigma$. It is the part of $S^{inv}$ that is
\emph{linear}   in the moduli. As such,  $\hat{S}$  is a proper
`special coordinate', i.e., a scalar field of an $N=2$ vector
multiplet, a property not shared by the full invariant dilaton
$S^{inv}$.}
\begin{equation}\label{Shat}
\hat{S}:=S-\frac{1}{4\pi}T_{-}\ .
\end{equation}
The two singlet scalars are therefore chosen to be
$(\hat{S},T_{+})$ .

We assume that $(C^a,\hat{S},T_+)$ are special coordinates of a
symplectic section
%$(X^I,F_{J}$ $(I,J=0,1,\ldots,5)$
 for which a holomorphic prepotential
-- denoted by $\cF_{\rm in}(C^a,\hat{S},T_+)$ --
exists.\footnote{This assumption is supported by the tree level
approximation discussed in the previous section and the
self-consistency of our one-loop result (Sections \ref{Sloop} and
\ref{qsymm}). Using   the tree level approximation, however, one
can also show
  that the rank two gauge groups
at $T=U=1$ and $T=U=\rho$ (which are beyond the scope of the
present paper) can\emph{not} be manifestly realized in a
symplectic basis with a prepotential.} The relation to the
notation of  Appendix A is  given by
\begin{equation}\label{loopvariables}
\frac{X^j}{X^{0}}=t^j=(iC^a,i\hat{S},iT_{+}), \qquad j=1, \ldots,
5\ .
\end{equation}

Thus far, the only difference to the tree level case is that  the
singlet $t^4$ is to be identified with $i\hat{S}$ instead of $iS$.
It is therefore not surprising that many of the general
conclusions we drew in Section \ref{Stree}  go through   for   the
one-loop corrected theory as well. In particular, the $SU(2)$
transformation properties of the $X^{I}$ remain formally the same,
\begin{eqnarray}
\delta X^{a}&=&\Lambda^b \epsilon_{bca}X^{c}\nonumber\\
\delta X^{0}&=&\delta X^{4}\, =\, \delta X^{5}\, =\,
0,\label{SU2trafo2}
\end{eqnarray}
which, together with  the analogue of eqs. (\ref{Killingtrafo}),
implies that the Killing vectors do not get
renormalized:
\begin{equation}\label{Killingloop}
k_a^b=\epsilon_{abc}C^c, \qquad k_{a}^{+}=k_{a}^{\hat{S}}=0
\end{equation}

Similarly, one can repeat large parts of the analysis of the
scalar potential,
\begin{equation}\label{Vpot}
  V = 2 \ e^K (X^I k_I^{\bar\imath}) g_{\bar \imath j}\, (\bar X^J
k_J^j) \ .
 \end{equation}
Because of the manifestly positive semidefinite form, the  vacua
are again given by  $ \langle    \bar{X}^{J}k_{J}^{j} \rangle
  =0$,
which, in the light of (\ref{Killingloop}), again implies that any
Minkowski vacuum can be brought to the form $\langle C^1  \rangle
= \langle C^2 \rangle = 0 $ by means of an $SU(2)$ rotation. These
vacua are also the supersymmetric ones, and we see that the
one-loop corrections might change the shape of the scalar
poptential, but not its ground states.

So far, everything we `derived' in this section was completely
independent
 of the prepotential $\mathcal{F}_{\rm in}$ and solely based on the assumed
$SU(2)$ gauge invariance and the underlying $N=2$ supersymmetry.
The $SU(2)$ symmetry, however, also restricts the possible form of
the prepotential, which in turn allows us to make further
statements about the theory without the  detailed  knowledge of
the prepotential.

More precisely, the $SU(2)$ gauge symmetry of the theory requires
that $\cF_{\rm in}(C^a,\hat{S},T_{+})$ be $SU(2)$ invariant.
Consequently, the triplet $C^a$ can only appear via the $SU(2)$
invariant combination $(C^a C^a)$ (or  powers thereof). Hence, the
prepotential  has to be of the general form
\begin{equation}\label{Fisasum}
\cF_{\rm in}(C^a,\hat{S},T_{+})=\sum_{n=0}^{\infty}
H_{n}(\hat{S},T_{+}) (C^a C^a )^n,
\end{equation}
where $H_n(\hat{S},T_{+})$ denotes a set of as yet undetermined
functions of the singlets $\hat{S}$ and $T_{+}$. Determining these
functions  will be the content of Sections \ref{Floop},
\ref{qsymm} and \ref{largeR}, but a number of statements already
follow from the general form (\ref{Fisasum}). As an example, let
us again  consider the masses of the scalar fields.

First note that in a vacuum with
 $\langle C^1  \rangle = \langle C^2 \rangle = 0 $, the metric components
$g_{1\bar{j}}$ and $g_{2\bar{j}}$ simplify to
\begin{eqnarray}\label{g11g22}
g_{1\bar{j}}&=&\delta_{1 j}
 e^K[(\mathcal{F}_{\rm in})_{11}+(\bar{\mathcal{F}}_{\rm in})_{11}]\ ,\nonumber\\
g_{2\bar{j}}&=&\delta_{2 j}
 e^K[(\mathcal{F}_{\rm in})_{11}+(\bar{\mathcal{F}}_{\rm in})_{11}]\ ,
\end{eqnarray}
where $\langle (\mathcal{F}_{\rm in })_{11} \rangle = \langle
(\mathcal{F}_{\rm in })_{22} \rangle$ has been used. A closer
inspection of (\ref{Vpot}) then reveals that the only
non-vanishing second derivatives of $V$ are
\begin{eqnarray}
\partial_{1}\partial_{1}
V\Big|_{ \langle C^1 \rangle =  \langle C^2 \rangle  =0}    &=&
-4 e^{K} g_{\bar{2}2} ({\bar{C}}^{3})^2   \ , \nonumber\\
\partial_{1}\partial_{\bar{1}}
V\Big|_{ \langle C^1 \rangle =  \langle C^2 \rangle  =0}    &=&
4 e^{K} g_{\bar{2}2} |C^{3}|^2   \ , \nonumber\\
\partial_{\bar{1}}\partial_{\bar{1}}
V\Big|_{     \langle C^1 \rangle =  \langle C^2 \rangle  =0}
        &=&- 4 e^{K} g_{\bar{2}2}(C^{3})^2\
\end{eqnarray}
and analogously for the derivatives with respect to $C^2$,
$\bar{C}^2$ (remembering $\langle g_{\bar{2}2} \rangle = \langle
g_{\bar{1}1} \rangle$). These mass matrices are again diagonalized
by a decomposition as in (\ref{decompo}). In terms of the corresponding
fields $a^{1,2}$ and $b^{1,2}$, the only non-vanishing derivatives are then
\begin{equation}
\frac{\partial^2 V}{\partial b^1 \partial b^1} \Big|_{     \langle
C^1 \rangle =  \langle C^2 \rangle  =0} =\frac{\partial^2
V}{\partial b^2 \partial b^2} \Big|_{     \langle C^1 \rangle =
\langle C^2 \rangle  =0} = 16e^Kg_{\bar{2}2} |C^3|^4.
\end{equation}
Taking into account the corresponding kinetic terms,
\begin{eqnarray}
\mathcal{L}_{\rm kin}&=& -g_{1\bar{1}}\partial_{\mu}C^1
\partial^{\mu} \bar{C^1} -
g_{2\bar{2}}\partial_{\mu}C^2 \partial^{\mu} \bar{C^2}\nonumber\\
&=&\frac{1}{2}(2g_{\bar{1}1}|C^3|^2)(\partial_{\mu}a^1
\partial^{\mu} a^1+
\partial_{\mu}b^1 \partial^{\mu} b^1+
\partial_{\mu}a^2 \partial^{\mu} a^2+\partial_{\mu}b^2 \partial^{\mu} b^2),
\end{eqnarray}
 one obtains for
the masses of
 the corresponding canonically normalized scalar fields
\begin{equation}\label{scalarmass2}
M^2=8e^{K} |C^3|^2=8e^{K}|T_{-}|^{2}.
\end{equation}
This has the same form as in the tree level approximation, but the
non-holomorphic K\"{a}hler potential $K$ now contains quantum
corrections. The \emph{holomorphic} mass, $m$, however, is the
same as it was at tree level,
\begin{equation}\label{holomorphicmscalar2}
m= T_{-},
\end{equation}
as anticipated in section~\ref{Stree}.

As the vacua with   $\langle C^1  \rangle = \langle C^2 \rangle =
0 $ are $N=2$ supersymmetric, we  expect the vector fields
$A_{\mu}^{1,2}$ (i.e., the $W^{\pm}$ bosons) to acquire the same
mass (\ref{scalarmass2}) as their scalar superpartners by
absorbing the massless fields $a^{1,2}$ in a supersymmetric Higgs
effect. This is again easy to verify: Just as in the tree level
case, the mass term for the vector fields arises from the square
of the covariant derivative \beq D_\mu C^a =
\partial_\mu C^a - k^{a}_b A_\mu^b \, \eeq which leads to the mass term
\begin{equation}
-g_{2\bar{2}}|C^{3}|^2 (A_{\mu}^{1})^2 -g_{1\bar{1}}|C^{3}|^2
(A_{\mu}^{2})^2.
\end{equation}
The corresponding kinetic terms are (see eq.\ (\ref{agsg}))
\begin{equation}
\frac{1}{4}\frac{(\bar{\mathcal{F}}_{11}+\mathcal{F}_{11})}{4}
F_{\mu\nu}^{1}F^{\mu\nu 1} +
\frac{1}{4}\frac{(\bar{\mathcal{F}}_{22}+\mathcal{F}_{22})}{4}
F_{\mu\nu}^{2}F^{\mu\nu 2}
\end{equation}
so that, remembering (\ref{g11g22}),
     one  indeed   obtains the same mass matrices as above
\begin{eqnarray}
M^2_{ab}  &=& 8 e^K\ {\rm diag} (|C^3|^2,|C^3|^2,0)
= 8 e^K\ |T_-|^2\ {\rm diag} (1,1,0) \label{MW}\nonumber\\
m_{ab}&=&   T_- \ {\rm diag} (1,1,0) \label{holomorphicm}
\end{eqnarray}

%%%%%%%%%%%%%%%%%%%%%%%%%%%%%%%%%%%%%%%%%%%%%%%%%%%%%%%%%%%
\subsection{Determining $\mathcal{F}_{\rm in}$}\label{Floop}
We are now ready to determine the complete perturbative
prepotential $\mathcal{F}_{{\rm in}}(C^{a},\hat{S},T_+)$ that
encodes the non-singular effective action $S_{\rm
in}[C^{a},\hat{S},T_+]$. The defining property of $S_{\rm
in}[C^{a},\hat{S},T_+]$ is that integrating out $C^{1}$ and
$C^{2}$ and going over to the variables $(S,T_{+},T_{-})$ should
reproduce the singular   action $S[S,T_+,T_-]$ based on the
perturbative  prepotential
$\mathcal{F}(S,T_{+},T_{-})=\mathcal{F}^{(0)}+\mathcal{F}^{(1)}$
given  in Section \ref{STUreview}.

As we are now going beyond tree level,    integrating out $C^{1}$
and $C^{2}$ is no longer equivalent to simply setting these fields
equal to zero. Instead, one now also has to take into account
threshold effects that arise from Feynman
 diagrams in which $C^{1}$ and $C^{2}$ (or their
superpartners) run in loops.

In practice this means that $\mathcal{F}(S,T_{+},T_{-})$ is
obtained from  $\mathcal{F}_{{\rm in}}(C^{a},\hat{S},T_+)$ in a
two-step process  (see also \cite{MZ}): First one sets
$C^{1}=C^{2}=0$ in $\mathcal{F}_{{\rm in}}$. This will then yield
an auxiliary prepotential $\mathcal{F}_{{\rm in}}^{{\rm
truncated}}\equiv \mathcal{F}_{{\rm in}}\big|_{C^{1}=C^{2}=0}$
  which only depends on $(C^3,\hat{S},T_{+})$ (or, alternatively, on
  $(S,T_{+},T_{-})$). If there were no threshold effects,
  this would already be the prepotential $\mathcal{F}$ in which the
$W^{\pm}$ bosons have been integrated out. If threshold effects do
exist,
 however,
$\mathcal{F}_{{\rm in}}^{\rm truncated}$ and $\mathcal{F}$ will
differ by an additional term $\delta \mathcal{F}$ which subsumes
all effective interactions that are generated by diagrams with
$C^{1}$ and $C^{2}$ running in loops, i.e., one has
\begin{equation}\label{Fsum}
\mathcal{F}=\mathcal{F}_{{\rm in}}^{{\rm truncated}} + \delta
\mathcal{F}.
\end{equation}

In our case, the threshold corrections introduce a logarithmic
dependence on the holomorphic mass of the $W^\pm$ gauge bosons
into the (Wilsonian) gauge couplings $g_W$ \cite{weinberg,KL,DKLL}
\beq \delta g_W^{-2} = - \ft{b}{16\pi^2} \log |m|^2\ , \eeq where
$b$ is the one-loop coefficient of the $\beta$-function. The
definition of the Wilsonian gauge coupling is exactly as in $N=1$
supergravity where $g_W$ is determined by a holomorphic function
\cite{KL}. In $N=2$ supergravity $g_W^{-2}$ is determined by the
matrix of second derivatives of $\cF$ \cite{DKLL} and for the case
at hand we find% 
\footnote{Note that the non-holomorphic piece in the definition
(\ref{Ndef}) of $\mathcal{N}_{IJ}$ does not contribute to the harmonic 
Wilsonian gauge couplings.}
\beq\label{gW} \delta g_W^{-2} =    \frac14
(\partial_-^2 \delta\cF + \bar\partial_-^2 \delta\bar\cF) =
\ft{1}{4\pi^2} \log |m|^2\ , \eeq where we used $b_{SU(2)}=-4$. Note
that the definition of $m$ includes the choice of a
(field-independent) cut-off scale which in supergravity has to be
proportional to the Planck mass. This in turn implies that the
right hand side of (\ref{gW}) is defined only up an arbitrary
additive constant. Using (\ref{holomorphicmscalar2}) this implies
\begin{equation}\label{deltaF}
\delta \mathcal{F}= \ft{1}{2\pi^2}\ T_-^2\log T_- + \A T_{-}^{2} +
\B(T_{+})T_{-} + \C(T_{+}),
\end{equation}
where $\A$ is the arbitrary constant while $\B(T_{+}), \C(T_{+})$
are a priori undetermined functions of $T_+$. The prepotentials
with and without the $W^{\pm}$ bosons are thus related by
\beq\label{diff} \cF = \cF^{\rm truncated}_{\rm in}
 + \ft{1}{2\pi^2}\ T_-^2\log T_-
+   \A T_{-}^{2} + \B(T_{+})T_{-} + \C(T_{+}). \eeq

As we see, integrating out $W^\pm$ introduces a logarithmic
singularity into $\cF$ while $\cF_{\rm in}$ has to be
non-singular. Thus we can now go backwards and compute $\cF^{\rm
truncated}_{\rm in}$
 by first
subtracting the logarithmic divergence $\delta\mathcal{F}$   from
$\cF$, which is
   given by
(\ref{WeylChamber1}) and (\ref{Fst}). $\cF_{\rm in}$ is then
obtained from $\cF^{\rm truncated}_{\rm in}$ by replacing every
$T_{-}^{2}$ by $C^{a}C^{a}$. For this to be possible, $T_{-}$
should appear in $\cF^{\rm truncated}_{\rm in}$ only in terms of
\emph{even} powers (cf. (\ref{Fisasum})). We will see whether this
is indeed the case.

The first step is therefore to expand $\cF$ near $T=U$ in order to
isolate the logarithmic singularity. The last term in
(\ref{WeylChamber1}) is manifestly non-singular in this limit
while the second term can be expanded using (\ref{PolyLogOne}). In
the region $\textrm{Re} T_{-} > 0$  one finds
 \bea\label{Zwischen}
\cF &=& -S(T_{+}^{2}-T_{-}^{2})-\ft1{12\pi} (T_{+}-T_{-})^3 -
 \ft{1}{(2 \pi)^4} \zeta(3) + \ft1{24\pi} T_- -
\ft{3}{4 \pi^2} T_-^2  - \ft1{3\pi} T_-^3 + O(T_-^4) \nonumber \\
 & & + \ft1{2\pi^2} T_-^2 \log(4\pi T_-)
 - \ft{1}{(2\pi)^4}\sum_{k,l=0}^{\infty} c_1(kl) Li_3
\left( e^{-2 \pi [(k+l)T_{+} + (k-l) T_{-}]} \right) \ . \eea In
appendix B (eqs.\ (\ref{PolyLogOne}), (\ref{even})) we show that
the terms denoted by $ O(T_-^4)$ involve  at most \emph{even}
powers $(T_{-})^{2n}$ with $n\geq 2$ The last term in
(\ref{Zwischen}) is analytic near $T_{-}=0$ and manifestly
invariant under $T_{-}\rightarrow -T_{-}$. Viewed as a power
series in $T_{-}$, it therefore also contains only \emph{even}
powers of $T_{-}$. Furthermore, the only non-analytic piece is the
logarithmic term
\begin{equation}
\ft1{2\pi^2} T_-^2 \log(4\pi T_-).
\end{equation}
Remembering (\ref{diff}), we see that in $\mathcal{F}_{{\rm
in}}^{{\rm truncated}}$  this term is precisely
  canceled
by the one-loop   threshold correction $\delta \mathcal{F}$. Thus,
as desired, $\mathcal{F}_{{\rm in}}^{{\rm truncated}}$ is
 analytic
  near $T_{-}=0$ and reads
\begin{eqnarray}\label{Ftrunc}
\cF_{{\rm in}}^{{\rm truncated}}
 &=& -S(T_{+}^{2}-T_{-}^{2})-\ft1{12\pi} (T_{+}-T_{-})^3 -
 \ft{1}{(2 \pi)^4} \zeta(3) + \ft1{24\pi} T_- 
  - \ft1{3\pi} T_-^3 \nonumber\\
& & + O(T_-^4)
  - \ft{1}{(2\pi)^4}\sum_{k,l=0}^{\infty} c_1(kl) Li_3
\left( e^{-2 \pi [(k+l)T_{+} + (k-l) T_{-}]}
\right) \nonumber\\
& & - [\A + \ft{3}{4 \pi^2} -\ft{1}{2\pi^2}\log (4\pi)]
 T_{-}^{2} - \B(T_{+})T_{-} - \C(T_{+})  \ .
\end{eqnarray}
As mentioned above, the full prepotential $\mathcal{F}_{\rm in}$
is now obtained from $\cF_{{\rm in}}^{{\rm truncated}}$  by
reversing the truncation of the $W^{\pm}$ bosons. At tree level,
this was done by simply promoting every $T_{-}^{2}$ to the $SU(2)$
invariant combination $C^{a}C^{a}$. However, a   closer inspection
of (\ref{Ftrunc})
 reveals that
this is not possible here,  because there are \emph{cubic}
  powers of $T_{-}$ which cannot
cancel against any other term (there are also linear terms in
$T_{-}$, but we will discuss them later).

The source of this problem is of course that we are still working
with the variables   $(S,T_{+},T_-)$ that were suitable for the
tree level approximation. As explained in Section \ref{Sloopgen},
the loop corrected version instead requires  working with  the
quantities $(\hat{S},T_+,T_-)$ in terms of which  the Weyl twist
$\sigma$ becomes diagonal. Only in
 terms of the variables $(\hat{S},T_+,T_-)$ should one expect the prepotential
$\mathcal{F}_{\rm in}^{\rm truncated}$ to be even in $T_{-}$. And
indeed, inserting (\ref{Shat}) into (\ref{Ftrunc}), one obtains
\begin{eqnarray}\label{Ftrunc2}
\cF_{{\rm in}}^{{\rm truncated}}
 &=& -\hat{S}(T_{+}^{2}-T_{-}^{2})-\ft1{3\pi} T_{+}^{3} +\ft1{4\pi}
T_{+}(T_{+}^{2}-T_{-}^{2}) - \ft{1}{(2 \pi)^4} \zeta(3) +
\ft1{24\pi} T_-  + O(T_-^4) \nonumber\\
& &
  - \ft{1}{(2\pi)^4}\sum_{k,l=0}^{\infty} c_1(kl) Li_3
\left( e^{-2 \pi [(k+l)T_{+} + (k-l) T_{-}]}
\right)\nonumber\\
&&  - [\A + \ft{3}{4 \pi^2} -\ft1{2\pi^2}\log(4\pi)] T_{-}^{2} -
\B(T_{+})T_{-} - \C(T_{+})  \ .
\end{eqnarray}
We see that the disturbing cubic terms in $T_{-}$ have indeed
disappeared.

Let us now turn  to the terms quadratic in $T_-$. As we discussed
above, the constant $\A$ is undetermined by the subtraction
procedure, and so one is   free to choose $\A =-\ft{3}{4 \pi^2}
+\ft1{2\pi^2}\log(4\pi)$ in order to simplify eq.\ (\ref{Ftrunc2}).  

The  linear term
\begin{equation}\label{linear}
\ft1{24\pi} T_-\ ,
\end{equation}
on the other hand, only leads to a constant shift in one of the numerous theta
angles. Such a term can always be neglected, because it is part of
the ambiguity \cite{CFG,DKLL,HM} in the prepotential $\mathcal{F}$
we have mentioned below eq.\ (\ref{WeylChamber1}). The same is true
for the real part of a possible  constant term in $\B(T_{+})$ as
well as for a possible linear term in $\B(T_{+})$ with imaginary
coefficient. All other terms in $\B(T_{+})$, however, have to
vanish from the outset in
  order for $T_{-}$ to appear only with even powers. Modulo
irrelevant changes in theta angles, we have  thus derived
\begin{equation}\label{B=0}
\B(T_{+})\equiv 0\ .
\end{equation}
The full prepotential $\mathcal{F}_{\rm in}$ is then obtained by
simply replacing every $T_{-}^{2}$ in (\ref{Ftrunc2}) by
$C^{a}C^{a}$. 

It remains to determine 
the unknown function $\C(T_{+})$.
In principle, 
this could be done in a similar way as in our discussion following eq.\  
(\ref{Fsum}) by considering the couplings $ \mathcal{F}_{++}$, 
$\mathcal{F}_{+0}$
and $\mathcal{F}_{00}$. As the two multiplets we integrate out are not charged
with respect to the corresponding vector fields $A_{\mu}^{+}$ and 
$A_{\mu}^{0}$, the gauge couplings of the latter should not feel the shift 
$\delta \mathcal{F}$  and remain unchanged in the integrating out process.
This would suggest $A_{0}(T_{+})\equiv 0$. In   
the following two 
sections, we will see that this expectation is supported 
by a completely independent line of argument. As we will show, 
$A_{0}(T_{+})$ is
already strongly constrained
by the quantum symmetry and the proper large radius limit.

%%%%%%%%%%%%%%%%%%%%%%%%%%%%%%%%%%%%%%%%%%%%%%%%%%%%%%%%%%%%%%%%%%%%%%%
\section{Quantum symmetries of $\cF_{\rm in}$}
\setcounter{equation}{0}\label{qsymm}

At $T_-=0$ the original $\SLtwo_{T}\times \SLtwo_{U}$ quantum
symmetry reduces to the diagonal $\SLtwo_{+}$ acting on $T_+$ and
$T_{-}$ as \beq\label{SLdiag} T_+\to {aT_+ -ib\over icT_++d}\ ,
\qquad T_- \to T_-\ . \eeq This symmetry should be respected by
$\cF_{\rm in} $, which we explicitly check in this section. In
addition, this consistency check will confirm that $\A$ is a
constant and further constrain  $\C(T_{+})$ in eq.\
(\ref{Ftrunc2}).

The generic form of $\cF_{\rm in}$ is given in (\ref{Fisasum}),
which,  near $T_- =0$, can be approximated by the first two terms
\beq \label{Fexp} \cF_{\rm in}(\hat{S},T_+,C^a) = H_0(\hat S, T_+)
+ H_1(\hat S, T_+) C^aC^a + {\cal O}(C^4)\ . \eeq Since we have
already determined the tree level contribution to these functions
in (\ref{Ftree}), we can parameterize $H_0$ and $H_1$ more
conveniently by \beq
 H_0(\hat S, T_+) =  -\hat{S}T_{+}^{2}+ h(T_+)\ ,\qquad
H_1(\hat S, T_+) = \hat{S} + f(T_+)\ , \eeq where $h$ can be viewed
as the loop corrections to the K\"ahler potential of $T_+$ (also
known as the four-dimensional Green-Schwarz term
\cite{DKLL,AFGNT}), while $f(T_+)$ is the one-loop correction to
the $SU(2)$ gauge coupling. As we will show, appropriate
derivatives of $h$ and $f$ transform as modular forms which can be
computed from (\ref{hfinal1}).

Let us first focus on $h(T_+)$. For this coupling there is a
closely related computation we can make use of. In ref.\
\cite{KLT} the two-parameter $ST$-model was investigated. Its
prepotential including one-loop corrections is given by $\cF =
-ST^2 + h(T)$ with an $\SLtwo$ acting on $T$.
% and a logarithmic singularity at $T=1$.
Using arguments outlined in \cite{DKLL}, it was shown in
\cite{KLT} that $\partial_T^5 h(T)$ has to be a modular form of
weight $+6$. The exact same arguments can be used here to conclude
that  $\partial_+^5 h(T_+)$ has to be a modular form of weight
$+6$. Furthermore, in the $ST$-model the second derivative
$\partial_T^2 h$ has a logarithmic singularity at $T=1$ which
arises from the fact that additional massless states appear at
$T=1$ which lead to a gauge symmetry enhancement. However, in our
case the second derivative $\partial_+^2 h(T_+)$ has a logarithmic
singularity both at $T_+=1$ and $T_+= \rho$, since charged states
become massless at both  points. The coefficient of the
singularity is set by the $\beta$-function of the gauge group
opening up. More precisely, one has \beq
\partial_+^2 h(T_+) = + {b_{SU(2)}\over 4\pi^2} \log(T_+ -1) +
{b_{SU(3)}\over 4\pi^2} \log(T_+ -\rho)+ \textrm{finite}\ . \eeq
Using $b_{SU(2)} = -4$ and $b_{SU(3)}=-6$ this implies
 \beq\label{sing}
\partial_+^5 h(T_+) = -{2\over \pi^2} {1\over (T_+ -1)^3}
- {3\over \pi^2} {1\over (T_+ -\rho)^3} + \textrm{finite}\ . \eeq

In order to check the above singularity structure and the modular
properties,
 we will now compute $\partial_+^5 h$ by relating it to the modular forms
(\ref{hfinal1}). Truncating out $C^{1},C^{2}$ from (\ref{Fexp})
gives
\begin{equation}\label{Fexptrunc}
\cF^{\rm truncated}_{\rm in}(\hat{S},T_+,T_{-}) =
-\hat{S}(T_{+}^{2}-T_{-}^{2})+h(T_+) + f(T_+)\ T_{-}^{2} + {\cal
O}(T_{-}^{4}) \ .
\end{equation}
Hence,
\begin{eqnarray}
h(T_{+})&=&\cF^{\rm truncated}_{\rm in}\big|_{T_{-}=\hat{S}=0}\nonumber\\
&=& (\cF-\delta \cF)\big|_{T_{-}=S=0}\nonumber\\
&=&(\cF^{(1)}-\delta \cF)\big|_{T_{-}=0},\label{hF}
\end{eqnarray}
where we have used (\ref{Fsum}) and (\ref{Shat}) in the second and
$\cF^{(0)}\big|_{S=0}=0$ in the  third line. Using the explicit
form (\ref{deltaF})  of $\delta \cF$, we then obtain
\begin{equation}\label{hFC}
 \partial_+^5 h=\partial_+^5\cF^{(1)}\big|_{T_{-}=0}-\partial_+^5 \C(T_{+}).
\end{equation}
In order to compute $ \partial_+^5\cF^{(1)}\big|_{T_{-}=0}$ it is
convenient to define \beq\label{Idef} I_\pm = (\partial_T^3 \pm
\partial_U^3) \cF^{(1)}\ , \eeq where the third derivatives of
$\cF^{(1)}$ are given in (\ref{hfinal1}). Expanding $I_\pm$ near
$T_-=0$ one has \bea
I_+ &=& a_0(T_+) + a_2(T_+)\, T_-^2 + O(T_-^4) \nonumber\\
I_- &=& a_{-1} T_-^{-1}  + a_1(T_+)\, T_- + O(T_-^3) \eea 
where\footnote{This and some of the following calculations have been performed using Maple.}
\bea\label{aexp}
a_{-1} &=& \frac{1}{4\pi^2}\ ,\nonumber\\
a_0 &=&
-\frac1{4\pi}\, E_2 -\frac1{4\pi}\,{E_4^{2} \over E_6}\ ,\\
a_1&=& \frac{23}{216}\,{ E_4} +\frac{1}8\,{E_2}^{2}+{\frac{1}4\, {
E_2 E_4^{2} \over E_6}
-\frac{4}{27}}\, {E_6^{2}\over E_4^{2}} \ ,\nonumber\\
a_2 &=& -\frac {19\pi}{432}}\, { E_2}\,{ E_4}+\frac {23\pi}{216}
 \,{ E_6}-\frac {19\pi}{432} \,{E_2^{3}
-\frac {19\pi}{144}\,{E_2^{2} E_4^{2} \over E_6}+ \frac
{\pi}{144}\,\frac { E_4^{3}}{ E_6} \nonumber\\
&&+\frac {4\pi}{27}}\,\frac {
E_6^{2}E_2}{ E_4^{2}}- \frac{\pi}{24}\,{\frac { E_4^{4} E_2}{
E_6^{2}}\ . \nonumber \eea
Expressing $\partial_+^5\cF^{(1)}\big|_{T_{-}=0}$ in terms of
derivatives of $I_{\pm}$, we  obtain after some straightforward
algebra \bea\label{hfinal}
 \partial_+^5\cF^{(1)}\big|_{T_{-}=0}
&=& \Big(4\partial_+^2 I_+ +\frac32 \partial_-^2 I_+ -
\frac92 \partial_-\partial_+ I_-\Big)\Big|_{T_-=0}\nonumber\\
&=& 4\partial_+^2 a_0 - \frac92 \partial_+ a_1 +3 a_2\\
&=& -2\pi\Big({E_4^6\over E_6^3} -\frac{23}{18} {E_4^3\over E_6}
+\frac{8}{18}{E_6^3\over E_4^3} -\frac16 E_6\Big)\nonumber\ , \eea
where the last equation used repeatedly the derivatives of modular
forms given in appendix~C. As expected $\partial_+^5\cF^{(1)}$ is
indeed  a modular form of weight $+6$. It also is closely related
to the corresponding quantity for the $ST$-model computed in
\cite{KLT} but differs in the structure of the singularities to
which we turn to now.

In (\ref{sing}) we determined the singularities of $\partial_+^5
h$ which differs from $\partial_+^5\cF^{(1)}\big|_{T_{-}=0}$ by
the so far unknown $\partial_+^5 \C(T_{+})$ (c.f.\ (\ref{hFC})).
However, as we are going to see shortly
$\partial_+^5\cF^{(1)}\big|_{T_{-}=0}$ has precisely the right
singularity structure and modular properties to be exactly
  \emph{equal}
to $\partial_+^5 h$ so that $\partial_+^5 \C(T_{+})$
 has to vanish identically.
First of all, it is easy to see that
$\partial_+^5\cF^{(1)}\big|_{T_{-}=0}$ does have a triple pole at
$T_+=1$ and $T_+=\rho$. Using (\ref{Eprop}) and expanding \bea
E_6(iT_+) &=& i E_6' (T_+-1) + \ldots \nonumber\\
E_4(iT_+) &=& i E_4' (T_+-\rho) + \ldots \eea we infer that near
$T_+ =1$ the leading singularity is \beq
\partial_+^5\cF^{(1)}\big|_{T_{-}=0} \to -2\pi i {E_4^6(i)\over
E_6^{\prime 3}(i) (T_+-1)^3} = -\frac2{\pi^2} \frac1{(T_+-1)^3}
\eeq which is indeed consistent with (\ref{sing}). Similarly, at
$T_+=\rho$ the leading singularity is \beq
\partial_+^5\cF^{(1)}\big|_{T_{-}=0}
\to -\frac{8\pi i}9 {E_6^3(\rho)\over E_4^{\prime 3}(\rho)
(T_+-\rho)^3} = -\frac3{\pi^2} \frac1{(T_+-\rho)^3}\ , \eeq again
consistent with (\ref{sing}). Finally, from the dual type IIA
vacua we know that for large $T_+$ the prepotential is at most a
cubic polynomial and hence \beq \lim_{T_+\to\infty} \partial_+^5 h
= 0 \ . \eeq Using (\ref{Enorm}) we indeed check \beq
\lim_{T_+\to\infty}
\partial_+^5\cF^{(1)}\big|_{T_{-}=0} =  2\pi(1-\frac{23}{18}
+\frac{8}{18} -\frac16) =0\ . \eeq
We thus conclude
\begin{equation}\label{C=0}
\partial_+^5 h= \partial_+^5\cF^{(1)}\big|_{T_{-}=0} \Longrightarrow
\partial_+^5 \C(T_{+})\equiv 0
\end{equation}
so that $\C(T_{+})$ can be at most a quartic polynomial in
$T_{+}$.

In a similar fashion we can compute $f(T_+)$. Using
(\ref{Fexptrunc}), (\ref{Fsum}), (\ref{Shat})
 and $\cF^{(0)}\big|_{S=0}=0$, one first
derives
\begin{eqnarray}
f(T_{+})&=&\frac{1}{2}[\partial^{2}_{-}
\cF^{\rm truncated}_{\rm in}]_{T_{-}=\hat{S}=0}\nonumber\\
&=& \frac{1}{2} [\partial^{2}_{-}(\cF-\delta \cF)]_{T_{-}=S=0}\nonumber\\
&=&\frac{1}{2}[\partial^{2}_{-}(\cF^{(1)}-\delta \cF)]_{T_{-}=0}\
,
\end{eqnarray}
so that
\begin{equation}\label{fFA}
\partial_{+}
f=
\frac{1}{2}\partial_{+}\partial^{2}_{-}\cF^{(1)}\big|_{T_{-}=0}\ .
\end{equation}
Furthermore, using (\ref{Idef}), (\ref{hF}) and (\ref{fFA}), one
easily verifies
\begin{eqnarray}
a_0\equiv I_+\Big|_{T_-=0}  &=&
\frac{1}{4}\left[\partial_{+}^{3}\cF^{(1)}+
3\partial_{+}\partial_{-}^{2}\cF^{(1)}\right]_{T_{-}=0}\nonumber\\
&=& \frac{1}{4}\left[\partial^{3}_{+} h +\partial^{3}_{+}\C\right]
+\frac{3}{2}\, \partial_{+}f\ .
\end{eqnarray}
Differentiating twice yields (remembering $\partial_{+}^5 \C=0$)
\beq
\partial_+^3 f = \frac23 \partial_+^2 a_0
-\frac16 \partial_+^5 h\ . \eeq
{}From (\ref{aexp}) and (\ref{hfinal}) and repeated use of
(\ref{Ederivatives}) we compute \beq\label{fresult}
\partial_+^3 f = -
\frac {\pi}{108}\, { E_2^3}+\frac {\pi}{4} \,E_2 E_4-\frac
{2\pi}{9} \,{E_6} -\frac {\pi}{36}\,{E_2^{2} E_4^{2} \over E_6}-
\frac {\pi}{6}\,\frac { E_4^{4}E_2}{ E_6^2} +\frac
{\pi}{36}\,\frac { E_4^{3}}{ E_6}+ \frac{4\pi}{27}\,\frac {
E_6^{3} }{ E_4^{3}} \ . \eeq As we see, this expression is not a
modular form and singular both at $T_+=1$ and at $T_+=\rho$. However,
as was stressed in ref.\ \cite{DKLL} at one-loop the dilaton $S$
transforms under modular transformations. Nevertheless, it is
possible to define a modular invariant, non-singular dilaton
$S^{inv}$ via \cite{DKLL} \beq \label{Sinvdef} S^{inv} = S
-\frac12\partial_T\partial_U\cF^{(1)} -\frac1{8\pi^2}
\log[j(iT)-j(iU)]\ . \eeq To see the modular properties of $f$ we
need to separate  $f$ into the part which is redefined into
$S^{inv}$ and the left over piece $f_{cov}$ defined by
\beq\label{fcov} \hat S + f = S^{inv} + f^{cov} \ . \eeq Using the
same strategy as before, we find
     \beq\label{Sind}
\partial_+^3 S^{inv}\Big|_{T_-=0} = -\frac18 \partial_+^5 h
+\frac14 \partial_+^3 f  -\frac1{8\pi^2} \partial_+^3
\log[\partial_+j]\ . \eeq Taking the third derivative of
(\ref{fcov}) evaluated at $T_-=0$ and inserting into (\ref{Sind})
we arrive at \beq
\partial_+^3 f^{cov} = \frac34 \partial_+^3 f +\frac18 \partial_+^5 h
+\frac1{8\pi^2} \partial_+^3 \log[\partial_+j]\ . \eeq Using
(\ref{hfinal}) and (\ref{fresult}) this finally gives
\beq\label{fcfinal}
\partial_+^3 f^{cov} = \frac1{4\pi^2} \partial_+^3\log E_4(T_+) \ .
\eeq
We see that for $f^{cov}$ the result considerably simplified
compared to (\ref{fresult}) and both the modular properties and
the singularity structure qualitatively changed. The right hand
side of (\ref{fcfinal}) is singular only at $T_+=\rho$
corresponding to the enhancement $U(1)\times SU(2)\to SU(3)$ as
expected. At $T_+=1$ on the other hand the enhancement is
$U(1)\times SU(2)\to SU(2)\times SU(2)$ and no charged states
contribute to the $SU(2)$ gauge couplings which is already present
at $T=U$. Thus $f^{cov}$ has to be finite at $T_+=1$ which is indeed
satisfied by the $\log E_4$-term.

Let us close this section by checking  the modular properties of
$f^{cov}$. {}From (\ref{Fexp}) we infer that $f(T_+)$ plays the
role of the one-loop corrections to the $SU(2)$ gauge coupling. In
$N=2$ supergravity the gauge couplings obey \cite{DKLL} \beq g^{-2}
= \R (\hat S+f) + {b\over 16\pi^2}\, K(S,T_+,C^a=0) \ , \eeq where
$K(S,T_+,C^a=0)$ is the tree level K\"ahler potential obtained
from (\ref{KSU2}) \beq \label{Ktransform} K= -\log (S+\bar S) -2
\log(T_++\bar T_+) \ . \eeq $K$  transforms under (\ref{SLdiag})
according to \beq K\to K +2 \log |icT_+ +d|^2\ . \eeq Since the
physical gauge couplings $g$ have to be modular invariant the
combination $\hat S+f = S^{inv} + f^{cov}$ has to compensate the
transformation (\ref{Ktransform}). Since $S^{inv}$ is modular
invariant by construction, the transformation law of $f^{cov}$ is
fixed to be \beq f^{cov}\to f^{cov} - {b\over 4\pi^2} \log (icT_+
+d) = f^{cov} + {1\over \pi^2} \log (icT_+ +d)\ , \eeq where the
last equation used $b_{SU(2)} =-4$. Thus we conclude
\beq\label{fcfinall}
 f^{cov} = \frac1{4\pi^2} \log E_4(T_+) \ 
\eeq
which is consistent with both (\ref{fcfinal}) and (\ref{Ktransform})
and fixed only up to an arbitrary constant, which can be identified with the
ambiguous constant $A_{2}$ in (\ref{deltaF}).

%%%%%%%%%%%%%%%%%%%%%%%%%%%%%%%%%%%%%%%%%%%%%%%%%%%%%%%%%%%%%%%%%%%%%%%
\section{The Large Radius Limit}\label{largeR}
\setcounter{equation}{0}

In this section, we perform the large radius limit of the theory
and show the consistency of our results with the results obtained
in ref.\ \cite{MZ}\footnote{In fact, the results of
\cite{MZ} were the motivation for the present analysis.} for $d=5$. In this
limit, one circle of the $T^2$ is decompactified, and one obtains
heterotic string theory on $K3 \times S^1$ \cite{AntFerTay}. The
low energy limit of this theory is five-dimensional, $N=2$
supergravity coupled to $n_V - 1$ vector multiplets and $n_H$
hypermultiplets, where $n_V,n_H$ count four-dimensional
supermultiplets. As before, the hypermultiplets can be
consistently ignored. The couplings of five-dimensional vector
multiplets to $N=2$ supergravity are encoded in a cubic
prepotential \cite{GST}. The vector multiplet moduli $T^i$ are
real, rather than complex, and the moduli space is a cubic
hypersurface, \beq {\cal V}(T^i) = \ft16 C_{ijk} T^i T^j T^k
\stackrel{!}{=} 1 \;, \label{5dPrep} \eeq where ${\cal V}(T^i)$ is
the five-dimensional prepotential, the $C_{ijk}$ denote a set of
constants,  and $i = 1, \ldots, n_V$. The underlying structure is
often referred to as  `very special geometry'  \cite{deWvP}.

Dimensional reduction of the five-dimensional supergravity theory
defined by (\ref{5dPrep}) over a circle of radius $R$ gives
four-dimensional $N=2$ supergravity with $n_V$ vector multiplets
and a `very special' (i.e., purely cubic) four-dimensional
prepotential: \beq {\cal F}(t^i) = \ft16 C_{ijk} t^i t^j t^k \;.
\label{PrepVS} \eeq Here, the $t^i$ are complex scalars whose real
parts are related to the  five-dimensional   scalars by \beq
\textrm{Re } t^i = R T^i.\label{rela}\eeq The imaginary parts of
the $t^{i}$ arise from the internal components of the
corresponding gauge fields. This means that in  a meaningful
decompactification limit the imaginary parts cannot have a vev,
and the $t^i$ have to be restricted to real values. In the rest of
this section, this will always be assumed, i.e., from now on $t^i$
stands for $\textrm{Re }t^i$, and inequalities such as $S>T>U$,
should be read as $\textrm{Re } S
>\textrm{Re }T>\textrm{Re }U$, etc.
With $t^i$ restricted to real values, (\ref{5dPrep}) and
(\ref{rela}) imply \beq {\cal F}(t^i) = \ft16 C_{ijk} t^i t^j t^k
=R^3 \;. \label{PrepVS2} \eeq

Whereas a five-dimensional prepotential must be purely cubic, a
four-dimensional prepotential is, in general, allowed to be an
arbitrary holomorphic function of the $t^i$ (possibly with
singularities on special loci). Therefore (\ref{PrepVS}) and
(\ref{PrepVS2}) only represent the pure supergravity contribution
that can be (and typically is) subject to further stringy
corrections:
 \beq {\cal F}(t^i) = \ft16 C_{ijk} t^i t^j t^k + \cdots = R^3 +
\cdots \;. \eeq If, however, such a four-dimensional prepotential
can be obtained by dimensional reduction from  five dimensions,
then these corrections must vanish in the decompactification limit
$R \rightarrow \infty$: \beq \lim_{R \rightarrow \infty} R^{-3}
{\cal F}(t^i) = {\cal V}(T^i) \;. \eeq

In order to make contact with \cite{MZ}, we need to switch to a
slightly different parameterization of the prepotential. This
corresponds, in the notation of \cite{CCLM}, to going from
`heterotic' to `type IIA' conventions: \beq S \rightarrow 4 \pi S
\;,\;\;\; {\cal F} \rightarrow -4 \pi {\cal F}\;. \eeq In this
convention, the prepotential without the $W^{\pm}$ bosons (eqs.
(\ref{Fst}), (\ref{WeylChamber1})) takes the form
\beq\label{neueVersion} {\cal F}_{>} = STU + \ft13 U^3 +
\frac{2}{(2\pi)^3} Li_3 \left( e^{-2 \pi (T-U)} \right) +
\frac{2}{(2\pi)^3} \sum_{k,l=0}^{\infty} c_1(kl) Li_3 \left( e^{-2
\pi (kT + lU)} \right) \;. \eeq We indicated by our notation that
this expression is valid in the Weyl chamber\footnote{As explained
in \cite{HM} the BPS states form an infinite dimensional Lie
algebra and one can generalize the notion of a Weyl chamber, which
is familiar from simple Lie algebras. In particular, $T>U$ and
$T<U$ define the two Weyl chambers of an $SU(2)$ subalgebra. The
corresponding group is the gauge group which is un-Higgsed at
$T=U$.} $S>T>U$. In order to perform the decompactification limit
inside this Weyl chamber we set $S = Rs$, etc. and take $R
\rightarrow \infty$, while keeping $s>t>u$ fixed . Using
(\ref{DecompLim}), we find \beq \lim_{R \rightarrow \infty} R^{-3}
{\cal F}(S,T,U) = {\cal V}(s,t,u) = stu + \ft13 u^3  \;,
\label{5dWeylChamber1} \eeq which is precisely the same
prepotential as one obtains directly in five-dimensional heterotic
string theory \cite{AntFerTay}.

The five-dimensional scalars
$s,t,u$ are subject to the hypersurface constraint (\ref{5dPrep}).
One can express them in terms of two unconstrained scalars. The
natural choice for these scalars are the five-dimensional
heterotic dilaton $\phi$, or, equivalently, the five-dimensional
heterotic string coupling $g_{(5)} = \sqrt{ {2 \pi}{/\phi}}$ and
the radius $r$ of the remaining circle. The relation between the
contrained scalars $s,t,u$ and the uncontrained scalars $\phi,r$
is \cite{AntFerTay}: \beq s= \frac{\phi}{2 \pi} - \frac{ \sqrt{2
\pi}}{3 \sqrt{\phi} r^3} \;,\;\;\; t = \frac{ \sqrt{2 \pi}
r}{\sqrt{\phi}} \;, \;\;\; u = \frac{ \sqrt{ 2 \pi}}{ \sqrt{\phi}
r } \;. \eeq The regime $s>t>u>0$ thus corresponds to $1<
\ft{r}{\sqrt{\alpha'}} < (2 \phi)^{3/2}$ \cite{AntFerTay}, i.e.,
the radius of the circle is larger than the self-dual radius
$\sqrt{\alpha'}$ and the heterotic string is weakly coupled.

Let us now take a different decompactification limit, where the
hierarchy between the moduli $T,U$ is reversed: $S>U>T$. The
four-dimensional prepotential in this region is
found by analytical continuation of (\ref{neueVersion}) using the
connection formula (\ref{ConFor}) for the polylogarithm: \bea
\cF_{<} &=& STU + \ft13 U^3 + \ft13 (T-U)^3 - \ft{i}{2}
(T-U)^2 - \ft16 (T-U)  \nonumber \\
 & & + \ft{2}{(2\pi)^3} Li_3 \left( e^{-2 \pi(U-T)} \right)
+ \ft{2}{(2\pi)^3} \sum_{k,l=0}^{\infty} c_1(kl) Li_3 \left( e^{-2
\pi (kT + lU)} \right) \;. \label{WeylChamber2} \eea 
Note that
(\ref{neueVersion}) and (\ref{WeylChamber2}) differ by polynomial
terms that come from the analytical continuation of $Li_3 \left(
e^{-2 \pi(T-U)} \right)$. The additional cubic term will survive
in the decompactification limit although the polylogarithm itself
goes to zero. Thus one of the polylogarithmic terms 
leaves a subtle shadow in the decompactification
limit. Note that it is precisely this term which 
is responsible for the fact that the
prepotentials in the two Weyl chambers are not just related by
exchanging $T$ and $U$. Just as the
non-trivial monodromy around $T=U$, this is caused by the
threshold corrections corresponding to the two charged vector
multiplets which become massless on this line.

The last term in (\ref{WeylChamber2}), which contains an infinite
number of further polylogarithmic terms,
is manifestly invariant under the exchange of 
$T$ and $U$ and does not contribute
to the monodromy of the prepotential around $T=U$.
It contains the contributions
of the infinitely many other BPS states of the heterotic string.
This term is non-universal, in the sense that it depends
on details of the BPS spectrum. For example, it will be different
for the closely related model with instanton numbers $(13,11)$. In
contrast, the first polylogarithmic term is universal in the sense
that it is fully determined by the fact that we have $SU(2)$
enhancement on the line $T=U$.

We can now take the decompactification limit $S>U>T \rightarrow
\infty$ and obtain 
\beq {\cal V}_{<} = stu + \ft13 u^3 + \ft13 (t-u)^3 = stu
+ \ft13 t^3 + (tu^2 - t^2 u)  \;, \label{5dWeylChamber2} \eeq valid
for $s>u>t$. Comparing (\ref{5dWeylChamber1}) and
(\ref{5dWeylChamber2}) we see that the two prepotentials differ by
the term $\ft13(t-u)^3$. This difference vanishes at $t=u$, so the
prepotential itself is a continuous function at $t=u$. The
resulting couplings in the Lagrangian, however, are discontinuous,
because they depend on derivatives of the prepotential. These
discontinuities in the couplings are the analogues of the
logarithmic branch cuts present in four dimensions.

Using five-dimensional field theory, one can show that
the difference $\ft13 (t-u)^3$ of the two prepotentials precisely
corresponds to the threshold corrections of two charged vector
multiplets \cite{Wit:96,IntMorSei,MZ}.
This shows that at $t=u$ the
$U(1)$ gauge group corresponding to the scalar $t-u$ is enhanced
to $SU(2)$. This result was in fact first found in
\cite{AntFerTay} by using perturbative heterotic string theory for
the compactification on $K3 \times S^1$. In terms of
five-dimensional heterotic variables, the second Weyl chamber
$s>u>t$ corresponds to a small radius $r < \sqrt{\alpha'}$ of the
circle, and one recognizes the the $SU(2)$ enhancement as the
usual $SU(2)$ gauge symmetry enhancement at the self-dual radius
$r=\sqrt{\alpha'}$ of the circle.

Using the field redefinition $s \rightarrow
s + u-t $, the prepotential (\ref{5dWeylChamber2}) takes the form
$stu + \ft13 t^3$ given in \cite{AntFerTay}. The fact that this
form actually involves a field redefinition is crucial for the
study of space-time geometries where the scalars evolve
dynamically from $t>u$ to $t<u$. Indeed, a naive  use of ${\cal
V}_{>} = stu + \ft13 u^3$ for $t>u$ and ${\cal V}_{<} = stu +
\ft13 t^3$ for $t<u$, with the same $s$ in both Weyl chambers,
leads to artificial space-time singularities, which are absent
when the correct continuation (\ref{5dWeylChamber2}) is used
\cite{KalMohShm,Moh:02}.

Let us now  consider a third decompactification limit, where we
keep $T_-=\ft12(T-U)$ small, so that we stay in the vicinity of
the enhancement locus. In this case, we should keep the charged
vector multiplets and work with the prepotential ${\cal
F}_{\mscr{in}}$ or, for simplicity, with the truncated version
${\cal F}^{\mscr{truncated}}_{\mscr{in}}$ (\ref{Ftrunc}). In  the
conventions used in this section,  ${\cal
F}^{\mscr{truncated}}_{\mscr{in}}$
  takes the following form\footnote{At this point, we have already used
$A_{1}(T_{+})\equiv 0$.}:
\bea \label{FtruncVar} \cF_{{\rm in}}^{{\rm truncated}}
 &=& S(T_{+}^{2}-T_{-}^{2}) + \ft1{3} (T_{+}-T_{-})^3 +
 \ft{2}{(2 \pi)^3} \zeta(3) - \ft1{6} T_-   + \ft4{3} T_-^3 \nonumber\\
& & + O(T_-^4)
  + \ft{2}{(2\pi)^3}\sum_{k,l=0}^{\infty} c_1(kl) Li_3
\left( e^{-2 \pi [(k+l)T_{+} + (k-l) T_{-}]}
\right) \nonumber\\
& & + 4 \pi \left( [\A +\ft{3}{4\pi^2} -\ft1{2\pi^2}\log(4\pi)]
T_{-}^{2} +  \C (T_{+}) \right) \ .
\end{eqnarray}

We will now show  that one gets a consistent decompactification
limit if one first takes $T_-$ to zero and then takes $S,T_+$ to
infinity. In order to keep track of the behaviour of the
prepotential away from the special locus $t_-=0$, we use that the
five-dimensional prepotential is purely cubic and perform the
limit at the level of the third derivatives: \beq \lim_{R
\rightarrow \infty} \lim_{T_- \rightarrow 0} \frac{ \der^3
{\cal F}^{\mscr{truncated}}_{\mscr{in}} } { \der t^i \der t^j \der
t^k } = \frac{ \der^3 {\cal V}^{\mscr{truncated}}_{\mscr{in}} } {
\der T^i \der T^j \der T^K } \;, \eeq where $t^i = S,T_+,T_-$ and
$T^i = s, t_+, t_-$.

We illustrate this by computing the term cubic in $t_-$. This term
is particularly important because it encodes the five-dimensional
  threshold corrections. From (\ref{FtruncVar}) we find:
\beq \lim_{R \rightarrow \infty} \lim_{T_- \rightarrow 0}
\frac{ \der^3 {\cal F} }{ \der T_-\der T_- \der T_- } = 6 \cdot (
-\ft13 + \ft43 ) = 6 \;. \eeq This implies \beq {\cal
V}_{\mscr{in}}^{\mscr{truncated}} = t_-^3 + \cdots \;, \eeq where
we used (\ref{PolyDiff}) and (\ref{PolySpecial}) to show that the
contribution of the polylogarithms vanishes in the limit.

Looking at the other third derivatives and  ignoring the    term $A_0(T_{+})$
for the moment,
 we find \beq {\cal
V}_{\mscr{in}}^{\mscr{truncated}} = s (t_+^2 - t_-^2) + \ft13 (t_+
- t_-)^3 + \ft43 t_-^3 \;. \label{5din} \eeq
To compare (\ref{5din}) with (\ref{5dWeylChamber1})
and (\ref{5dWeylChamber2}) we switch to the variables $s,t,u$: \beq
{\cal V}_{\mscr{in}}^{\mscr{truncated}} = stu + \ft13 u^3 + \ft16
(t-u)^3 = \ft12 ({\cal V}_{>} + {\cal V}_{<}) \label{5din1}\;. \eeq This is
precisely the truncated  five-dimensional prepotential derived in
\cite{MZ}. Just as in the four-dimensional case (section 4.2), a
 manifestly gauge invariant form is obtained by
introducing $\hat{s} = s-t_-$, which is the five-dimensional limit
of the Weyl-invariant dilaton $\hat{S} = S - T_-$(\ref{Shat}): \beq
{\cal V}_{\mscr{in}}^{\mscr{truncated}} = \hat{s} (t_+^2 - t_-^2)
+ \ft13 t_+^3 + t_+ t_-^2 \;. \label{5din2} \eeq In this basis all
fields   transform covariantly under the Weyl twist: $\hat{s}$ and
$t_+$ are invariant, and $t_-$ is mapped to $-t_-$. As $t_{-}$
only enters through the invariant $t_-^2$, the `untruncated'
prepotential is again obtained via a  substitution of the form
$t_-^2 \rightarrow  (c_1^2 + c_2^2 + c_3^2)$, where $c_i$
transform in the adjoint representation of $SU(2)$ \cite{MZ}.

We have thus shown that our decompactification limit of ${\cal
F}^{\mscr{truncated}}_{\mscr{in}}$ is consistent with the purely five-dimensional 
 result 
(\ref{5din1}) obtained in \cite{MZ} provided that 
the function $A_0(T_+)$ does not contribute to this limit, i.e., provided
that $\lim_{R\rightarrow \infty}\partial_{+}^{3}A_{0}(T_{+})=0$. 
From
Section \ref{qsymm}, we already  know 
that $A_{0}(T_{+})$ can be at most a quartic polynomial in 
$T_{+}$. The five-dimensional   decompactification limit now tells us that
$A_0(T_{+})$ can, in fact,  be at most a \emph{quadratic} polynomial:
$A_0(T_{+})=c_0+c_{1}T_{+}+c_{2}T_{+}^{2}$. As mentioned earlier,
these remaining terms are expected to vanish as well, because
they would give rise to changes in the gauge couplings 
of the spectator vector fields 
$A_{\mu}^{+}$ and $A_{\mu}^{0}$ when the $W^{\pm}$ multiplets 
are integrated out.

%%%%%%%%%%%%%%%%%%%%%%%%%%%%%%%%%%%%%%%%%%%%%%%%%%%%%%%%%%%%%%%%%%%%%%%
\section{Conclusions}\label{conclusions}
\setcounter{equation}{0}

In this paper, we have shown, in an explicit example, how to
determine a non-singular effective action near a singular subspace
of the moduli space of a string compactification. The key feature
of this effective action is that it includes modes that are
massive at a generic point in the moduli space but become massless
at the singularity. Starting from a singular effective action
where such modes have been integrated out, we carefully integrated
them back in and in this way derived an effective action valid in
the vicinity of the singularity, or, in other words, in a region of
 the moduli space where these modes are still light. Using a
combination of field-theoretical reasoning (the general structures
of $N=2$ Yang-Mills-supergravity actions and of threshold
corrections) together with some (in fact, little) knowledge of the
underlying microscopical string physics (only the type of the
additional massless states were needed) and symmetry arguments
(the residual T-duality at a fixed point) turned out to be
sufficient to determine the effective theory up to a few
irrelevant integration constants.

It was clear from the outset that such a non-singular effective
action should exist, but we find it interesting and useful to carry out
this derivation explicitly and determine a complete and consistent
description of the low energy physics near the singularity. Our 
calculations have exhibited many features which we expect to be 
generic. In particular, we have seen that although the prepotential
is a very complicated function, which involves an infinite number
of polylogarithmic functions, integrating in the charged vector
multiplets basically amounts to adding a term of the form
$\delta {\cal F} \simeq T_-^2 \log T_-$. This term is fixed
by pure field theory arguments, and is complelely
determined by the knowledge that
two charged vector multiplets become massless at $T_-=0$.
As we have seen, the resulting theory nevertheless 
has the  correct global
properties on the moduli space, i.e., it has the correct
singularities at the points $T_+=1,\rho$
of higher gauge symmetry enhancement
and exhibits the residual modular symmetry $SL(2,{\bf Z})_{+}$.
Modular symmetries are related to the presence of infinitely many
massive string states. What our results thus  demonstrate is  
that the field theory reasoning employed here
is able to capture such stringy properties.  
Also note that 
it is not completely obvious that (\ref{WeylChamber1}) does not contain
odd powers of $T_{-}$ apart from the first and the third. 
Since the infinitely many
massive modes do not interfere with the integrating in procedure 
we expect that the methods developed in this paper can be applied
to other cases as well.

%It was clear from the outset that such a non-singular effective
%action should exist, but we find it interesting and useful to carry out
%this derivation explicitly and determine a complete and consistent
%description of the low energy physics near the singularity. In
%particular, we saw that the non-universal terms in the
%prepotential, i.e. those terms which correspond to the other BPS
%states contributing to loop corrections, do not interfere with
%integrating in the two charged vector multiplets. Therefore we
%expect that the methods developed in this paper can be applied
%to other cases.
%our result is generic. 

%One might wonder what would be required in order to 
For example, it would be interesting to extend our
results to conifold points and conifold transitions in type II
string compactifications on Calabi-Yau threefolds. This
corresponds to situations where hypermultiplets
become massless, and we must distinguish between statements about the
vector multiplet sector and about the hypermultiplet sector. The
vector multiplet sector is still determined by its prepotential,
but now the singularities and monodromies of the generic prepotential
are not due to $SU(2)$
gauge symmetry enhancement, but to massless monopoles and dyons.
Integrating in these hypermultiplets must remove the non-trivial
monodromies of the prepotential around the conifold locus. 
Given the monodromies, one should be able to integrate in the
hypermultiplets in the same way as the vector multiplets considered
in this paper. Note that although the $SU(2)$ gauge symmetry is never
restored, it nevertheless leaves its imprint in the monopole and
dyon monodromies, as explained in \cite{CarLueMoh:95}.

It is much harder to say anything concrete about the
hypermultiplet sector, due to our lack of knowledge about generic
quaternionic manifolds. Whereas we can start in the vector
multiplet sector from a known prepotential, the metric on the
hypermultiplet moduli space is not known for the $STU$-model.
Therefore, any extension of our knowledge on this sector of the
theory is extremely valuable. One interesting question is the
structure of the metric and of the scalar potential in the
effective theory where the monopole hypermultiplets have been
integrated in. The particular structure of the scalar potential
corresponds to a non-generic gauging of the supergravity
Lagrangean (since one still has many flat directions) and requires
the hypermultiplet manifold to have specific isometries. It should
be interesting to investigate this in detail.

Another direction is the investigation of higher rank non-abelian
gauge groups in the perturbative heterotic string. This will
require a generalization of the present formalism, since, at least
in our example, these higher gauge symmetry enhancements cannot be
described in a basis for the symplectic section where a
prepotential exists. A reformulation purely in terms of sections
should also be useful for studying  non-Abelian gauge symmetry
enhancement in type II Calabi-Yau compactifications 
\cite{KatMorPle,KleEtAl}. 
Persuing this line of developement,
we expect to get a better understanding of the relation between
gauged supergravity, the geometry of Calabi-Yau manifolds and
M-theory.

%%%%%%%%%%%%%%%%%%%%%%%%%%%%%%%%%%%%%%%%%%%%%%%%%%
%%%%%%%%%%%%%%%%%%%%%%%%%%%%%%%%%%%%%%%%%%%%%%%%%%
%%
%%   Appendix
%%
%%%%%%%%%%%%%%%%%%%%%%%%%%%%%%%%%%%%%%%%%%%%%%%%%%
%%%%%%%%%%%%%%%%%%%%%%%%%%%%%%%%%%%%%%%%%%%%%%%%%%

\vspace{1cm}
\appendix
\noindent {\Large {\bf Appendix}}
\renewcommand{\theequation}{\Alph{section}.\arabic{equation}}

\setcounter{equation}{0}\setcounter{section}{0}

\section{$N=2$ gauged supergravity in $d=4$}\label{sugra}
In this appendix we collect the some facts about gauged $N=2$
supergravity in $d=4$ \cite{bw,wp,CDF,N=2}. A generic spectrum
contains the gravitational multiplet which contains the graviton
$g_{\mu\nu}, \mu,\nu =0,\ldots,3$ and the graviphoton $A_\mu^0$ as
bosonic components. In addition there can be $n_V$ vector
multiplets which feature $n_V$ vector bosons $A_\mu^i$ and $n_V$
complex scalars $t^i, i=1,\ldots,n_V$ as bosonic components.
Finally there are  $n_H$ hypermultiplets which contain $4n_H$ real
scalars $q^u, u=1,\ldots, 4n_H$. The bosonic part of the effective
action reads \cite{N=2}\footnote{Our  normalizations coincide with
 those of 
\cite{DKLL}.} 

\begin{equation}
  \label{agsg}
  S =  \int  \frac12 R - g_{i\bar\jmath} D_\mu t^i  D^\mu {\bar t}^j
- h_{uv} \partial_\mu q^u  \partial^\mu q^v
  + \frac{1}{8}\, \I \cN_{IJ} F^I_{\mu\nu} F^{J\mu\nu}
  + \frac{1}{4} \, \R \cN_{IJ} F^I \wedge F^J - V \ ,
\end{equation}
where $R$ is the Einstein term and $h_{uv}$ is the metric on a
quaternionic manifold, $\cM_H$, spanned by the scalars $q^u$ in the
hypermultiplets. As this part of the action is of no
importance for this paper we do not discuss them any further and
instead refer the reader to the literature \cite{N=2}. The metric
$g_{i\bar\jmath}$ is the metric on a special K\"ahler manifold,
$\cM_V$, spanned by the scalars $t^i$. Being special K\"ahler,
$g_{i\bar\jmath}$ can be derived from a  K\"ahler potential via
$g_{i\bar\jmath} =\partial_i \bar\partial_{\bar\jmath}K$, where
$K$ is not an arbitrary real function but determined in terms of a
holomorphic prepotential $F$ according to
\begin{equation}\label{Kspec}
K=-\log \Big[i \bar{X}^{I} (\bar t) F_{I}(X) - i X^{I}
(t)\bar{F}_{I}(\bar{X})\Big] \ .
\end{equation}
The $X^{I}, I=0,\ldots, n_V$ are $(n_V+1)$ holomorphic functions
of the  $t^i$. $F_{I}$ abbreviates the derivative, i.e.
$F_{I}\equiv \frac{\partial F(X)}{\partial X^{I}}$ and $F(X)$ is a
homogeneous function of $X^{I}$ of degree $2$, i.e.\ $X^{I}
F_{I}=2 F$. Using this homogeneity property one can go to special
coordinates defined by $X^0 = 1,  X^i = t^i$. In this
parameterization the K\"ahler potential can be written as
\begin{equation}
K= -\log\Big[2({\cal F}+ \bar{\cal F})-
           (t^i+\bar t^i)({\cal F}_i+\bar{\cal F}_i)\Big]\,,
\label{Kspecial}
\end{equation}
where ${\cal F} = i (X^0)^{-2} F(X)$.

$F^I_{\mu\nu}$ are the field strength of the gauge bosons where
$F^0_{\mu\nu}$ denotes the field strength of the graviphoton. The gauge
coupling functions $\cN$ are defined in terms of the prepotential
according to \beq \label{Ndef} {\cal N}_{IJ} = \bar F_{IJ} +2i\
\frac{\mbox{Im} F_{IK}\mbox{Im} F_{JL} X^K
 X^L}{\mbox{Im} F_{LK}  X^K X^L} \ .
\eeq

The covariant derivatives are given by
\begin{equation}\label{gaugeco}
{D}_\mu  t^i = \partial_\mu  t^i - k_I^i A_\mu^I\ ,
\end{equation}
where $k_I^i(t)$ are Killing vectors which generate isometries on
$\cM_V$\footnote{Of course it is also possible to gauge isometries
on the quaternionic manifold
 $\cM_H$ but since this does not occur in the present models
we do not discuss  this situation here.}
\begin{equation}\label{kdef}
\delta t^i \ = \ \Lambda^I k_I^i(t) \  .
\end{equation}
As a consequence of the Killing equation and the K\"ahler geometry
of $\cM_V$, the $k_I^i(t)$ are constrained to be holomorphic, i.e.\
$\bar\partial k_I^i(t)=0$ and furthermore  can be solved in terms
of Killing prepotentials $P_I$
\begin{equation}\label{kP}
k_I^i(t) = g^{i\bar j} \partial_{\bar j} P_I\ .
\end{equation}
The $P_I$ in turn are determined by \beq\label{PF} P_I = e^K( F_J
f^J_{IK} \bar X^K +  \bar F_J f^J_{IK} X^K )\ , \eeq where the
$f^J_{IK}$ are the structure constants of the symmetry group.
Finally, the potential is expressed in terms of the Killing
vectors and reads \beq\label{pot}
  V = 2 \ e^K X^I\bar X^J g_{\bar \imath j}\, k_I^{\bar\imath} k_J^j \ .
\eeq

%%%%%%%%%%%%%%%%%%%%%%%%%%%%%%%%%%%%%%%%%%%%%%%%%%%%%%%%%%%%%%%%%
\section{Polylogology}\label{polylog}
\setcounter{equation}{0}

In this appendix we assemble facts and useful formulae for
polylogarithmic functions, as they can be found, for example, in
refs.\ \cite{PruBryMar,HM,BCG}.

For $0<z<1$, the k-th  polylog is defined by the series expansion
\beq Li_k(z) = \sum_{n=1}^{\infty} \frac{z^n}{n^k}\ .
\label{PolyLogInfty} \eeq It can be continued to a multivalued
function on the complex plane. Polylogarithmic functions with
different values of $k$ are related by the equation \beq z
\frac{d}{dz} Li_k(z) = Li_{k-1}(z) \;. \label{PolyDiff} \eeq
Whereas the first polylog is related to the logarithm, \beq Li_1(z)
= - \log(1-z) \;, \eeq the polylogs with $k \leq 0$ are algebraic
functions: \beq Li_0(z) = \frac{z}{1-z} \;, \;\;\; Li_k(z) = \left(
z \frac{d}{dz} \right)^{-k} \frac{z}{1-z} \; \mbox{   for   }
k\leq -1 \;. \eeq From (\ref{PolyDiff}) one can derive integral
representations for the higher polylogs, $k \geq 1$, but we will
not need them. But in order to describe the behaviour of the
prepotential in the decompactification limit and on the
enhancement locus we need the following special
values:\footnote{$Li_{-k}(z)$ has a $k$-th order pole at $z=1$ for
$k>0$, whereas $Li_{0}(z)$ diverges logarithmically for $z
\rightarrow 1$.} \beq Li_k(0) = 0 \;,\;\;(\forall k \in {\bf Z})
\;\;\; \mbox{and}\;\;\; Li_k(1) = \zeta(k) \;,
\;\;\mbox{for}\;\;k>1. \label{PolySpecial} \eeq

The connection formula \cite{PruBryMar} relates the values at $z$
and $\ft1z$: \beq Li_k(z) + (-1)^k Li_k ( \ft1{z} ) = - \frac{(2
\pi i)^k}{k!} B_k \left( \frac{ \log(z) }{ 2 \pi i} \right) \,,
\;\;\mbox{for   } k > 0 \;, \eeq where $B_k( \cdot )$ are the
Bernoulli polynomials.\footnote{There is an analogous equation for
$k\leq 0$, where the right  hand   side is zero.} For $Li_3$ one
finds \beq Li_3(z) - Li_3(\ft1z) = - \ft16 \log^3(z) - \ft{i\pi}2
\log^2(z) + \ft{\pi^2}3 \log(z)\;.
%- \frac{1}{3!} \log^3(-z) - \frac{\pi^2}{6} \log(-z)\;.
\label{Contin3} \eeq For our purposes it is more natural to work
with the variable $x$, where $z = e^x$. In the main part of the
paper, $x$ is a modulus or a linear combination of moduli, and $x=0
\Leftrightarrow z=1$ corresponds to gauge symmetry enhancement,
while $x \rightarrow \infty \Leftrightarrow \ft1z \rightarrow 0$
corresponds to the decompactification limit. In terms of the
variable $x$, formula (\ref{Contin3}) becomes \cite{HM} \beq
Li_3(e^x) = Li_3 (e^{-x}) + \frac{\pi^2}{3} x - \frac{i \pi}{2}
x^2 - \frac{1}{6} x^3 \ . \label{ConFor} \eeq

The function $Li_3(e^{-x})$ has a logarithmic branch point at
$x=0$. Since this limit is relevant for the study of gauge
symmetry enhancement, it is useful to have an expansion of the
form \beq Li_3 (e^{-x}) \simeq  p(x) + q(x) \log(x) \mbox{   for
} x \rightarrow 0 \;, \label{PolyLogExpansion} \eeq where $p(x)$
and $q(x)$ are power series, \beq p(x) = \sum_{j=0}^{\infty} p_j
x^j \qquad \mbox{   and   }\qquad  q(x) = \sum_{j=0}^{\infty} q_j x^j \;. \eeq
This expansion can be analytically continued to an expansion for
$Li_3(e^x)$, using $\log(-x) = \log(x) + i \pi$. Plugging this
into the connection formula (\ref{ConFor}) and comparing term by
term one finds: \beq q_0 = q_1 =0 \;, \;\;\; q_2 = - \frac{1}{2}
\;, \;\;\, q_3 = q_4 = q_5 = \cdots = 0 \eeq and \beq p_1 = -
\frac{\pi^2}{6} \;, \;\;\; p_3 = \frac{1}{12} \;, \;\;\;p_5 = p_7
= p_9 = \cdots = 0 \;. \eeq The coefficients $p_{2i}$, $i=0,1,2,
\ldots$ can be obtained using (\ref{PolyLogInfty}). $p_0$ is fixed
by $Li_3(1) = \zeta(3)$ while the other coefficients can be found
by comparing derivatives of (\ref{PolyLogInfty}) with
(\ref{PolyLogExpansion}). In particular, the second derivative
fixes $p_2 = \ft34$. Combining all our results we have
 \cite{BCG,HM}\footnote{Our formula specifies some terms which were not
displayed in \cite{BCG,HM}. Ref.\ \cite{BCG} suggests the
existence of terms of the form $O(x^3)\log(x)$, but as we have
seen these are absent. Our formula is consistent with eq.\ (8.5)
of \cite{HM} after the change  of variables $x=-\log(1-y)$. We
thank G.\ Cardoso for discussions on this issue.}
\beq \label{PolyLogOne} Li_3(e^{-x}) \simeq  p(x) - \frac{1}{2}
x^2  \log(x) \ , \eeq where \beq\label{even}
 p(x) = \zeta(3) -
\frac{\pi^2}{6} x + \frac{3}{4} x^2 + \frac{1}{12} x^3
 + O(x^{2n}) \;, n= 2,3,4,\ldots
\eeq Note that the higher terms in $p(x)$ are even powers of $x$.
The odd powers, except the linear and the cubic term, are ruled
out by the connection formula. This is important for our
discussion of gauge symmetry.

To analyze the decompactification limit we need the first formula
of (\ref{PolySpecial}), or, being more precise about the
asymptotics, \beq Li_3 (e^{-x}) \simeq e^{-x}
\;,\;\;\;\mbox{for}\;\;\; x \rightarrow \infty \;.
\label{DecompLim} \eeq

%The fact that non-analytic terms in the expansion of $Li_3(e^x)$
%are given exactly by  $- \frac{1}{2}  x^2  \log(x)$ can also be
%deduced from the formula \cite{PruBryMar} \beq Li_n(z) - Li_n( e^{2
%\pi i} z) = \frac{2 \pi i}{\Gamma(n)} \log^{n-1} (z) \ , \eeq which
%computes the monodromy of $Li_n$ around $z=1$. Setting $z=e^{-x}$
%we find \beq Li_3(e^{-x}) - Li_3( e^{-x} e^{2 \pi i}) = \frac{2 \pi
%i}{\Gamma(3)} \log^2(e^{-x}) = i \pi x^2\ . \eeq If we now expand
%$Li_3(e^{-x})$ around $x =0$ according to \beq Li_3(e^{-x}) = p(x)
%+ q(x) \log(x) \eeq with power series $p(x),q(x)$, then the above
%monodromy must be reproduced by taking $x$ along a circle around
%$x=0$, i.e., we take $x \rightarrow e^{2 \pi i}x$ on the r.h.s.
%Only the $\log$ contributes to the monodromy: \beq p(x) - p(e^{2
%\pi i} x) + q(x) \log(x) - q(e^{2 \pi i}x) \log(x e^{2 \pi i}) = -
%q(x) 2 \pi i \eeq Comparing with the above monodromy formula for
%the polylog we find \beq - q(x) 2 \pi i = i \pi x^2 \Longrightarrow
%q(x) = - \ft12 x^2 \eeq This is consistent with what was deduced
%from the connection formula.

%%%%%%%%%%%%%%%%%%%%%%%%%%%%%%%%%%%%%%%%%%%%%%%%%%%%%%
\section{Modular Forms }\label{smodform}
\setcounter{equation}{0}

The modular group is defined by the following transformation:\footnote{%
It is common to choose a different convention for $T$ where real
and imaginary part are exchanged. More precisely, for $\tau = i T$
one has $\tau \to {a\tau +b\over c\tau + d}$.}
\beq\label{modtr} T \to {aT -ib\over icT + d}\ , \qquad ad - bc =
1\ , \qquad a,b,c,d \in {\bf Z} \;.\eeq On the fundamental domain of
this transformation there are two fixed points at $T=1$ and
$T=\rho\equiv e^{i\pi\over 6}$.

A modular form $E_k(iT)$ of weight $k$ is defined to be
holomorphic and to obey the transformation law \beq\label{modform}
E_k(iT) \to (icT+d)^k E_k(iT) \ . \label{mformdef} \eeq One can
show that there are no modular forms of weight 0 and 2, while at
weight 4 and 6 one has the Eisenstein functions \bea E_4 (q)&
\equiv &  1 + 240 \sum_{n=1}^\infty \frac{n^3 q^n}{1-q^n}
      \  =\ 1 + 240 q + 2160 q^2 \ldots\ ,\nonumber \\
E_6 (q) &\equiv&  1 - 504 \sum_{n=1}^\infty \frac{n^5 q^n}{1-q^n}
\ =\ 1 - 504 q -16632 q^2 \ldots\ , \label{Eisenstein} \eea where
$q\equiv e^{-2\pi T}$. From their definition one immediately
infers that they have been normalized such that \beq\label{Enorm}
\lim_{T\to\infty} E_4\ =\ 1\ =\ \lim_{T\to\infty} E_6\ . \eeq
Furthermore, both function have no pole on the fundamental domain
and $E_4$ has exactly one simple zero at $T=\rho$, while $E_6$ has
one simple zero at $T=1$
\beq\label{Eprop} E_4(i) \neq 0\ , \qquad
E_4(i\rho)= 0\ , \qquad E_6(i) = 0\ , \qquad E_6(i\rho)\neq 0\ .
\eeq
One can construct modular forms of arbitrary even weight from
products of these two Eisenstein functions.

A modular form which vanishes at $T=\infty$ is called a cusp form.
There is no cusp form of weight $r< 12$ and for $r=12$ there is
the unique cusp form $\eta^{24}$ where \beq \eta(q) \equiv
q^{1\over 24} \prod_{n=1}^\infty (1-q^n)\ , \qquad \eta^{24} =
{E_4^3 -E_6^2\over 1728} \label{etadef} \eeq is the Dedekind
$\eta$-function. ($\eta$ does not vanish at $\rho$ or $i$.)

One can also construct modular invariant functions but such a function
necessarily has a pole somewhere on the fundamental domain. The
$j$--function defined by \beq j(q)\ \equiv\  {E_4^3\over\eta^{24}}\
=\ {E_6^2\over \eta^{24}} +  1728 {E_4^3\over E_4^3 - E_6^2} =
q^{-1} + 744 + 196884 q + \ldots \label{jdef} \eeq has a simple
pole at $T=\infty$ and a triple zero at $T=\rho$.

Finally, the  Eisenstein series $E_2$ is defined by \beq E_2(iT) =
1 - 24 \sum_{n=1}^{\infty} {n q^n\over 1- q^n}\ . \eeq $E_2(iT)$ is
holomorphic, but not quite a modular form: \beq E_2 \to ( icT + d)^2
E_2 (iT) + \frac{6c}{\pi i}  ( icT + d ) \;. \eeq

The derivative of a modular form is in general not a modular form.
But using the transformation properties of $E_2$ one defines the
modular covariant derivative \beq Df(iT) := f'(iT) - k\;\frac{\pi
i}{6}  E_2(iT) f(iT) \ , \eeq where the prime denotes
differentiation with the respect to the argument, i.e.\  $f'(iT)
\equiv -i\partial_T f(iT)$. The covariant derivative maps
modular forms of degree $k$ to modular forms of degree $k+2$. Its
action on normalized Eisenstein series is \beq DE_k(iT)  = - k \;
\frac{\pi i}{6}  E_{k + 2}(iT) \eeq for $k=4,6, \ldots$. This can
be used to express derivatives of Eisenstein series in terms of
the Eisenstein series themselves. For $E_2'$ we also have a
relation: \beq E_2' - \frac{\pi i}{6} E_2 E_2 =  - \frac{\pi i}{6}
E_4 \;.\eeq

Also note that all higher Eisenstein series $E_k$, with
$k=8,10,12,\ldots$ are homogenous polynomials in $E_4,E_6$ (the
ring of modular forms is generated by $E_4,E_6$). For example: \beq
E_8 = E_4^2 \;, \;\;\;E_{10} = E_4 E_6 \;.\eeq In the text we need the
following derivatives: \bea\label{Ederivatives}
E_4' &=& \frac{2 \pi i}{3} ( E_4 E_2 - E_6 ) \;,\nonumber\\
E_6' &=& i \pi ( E_6 E_2 - E_8) = i \pi (E_6 E_2 - E_4^2) \;,\\
j' &=& - 2 \pi i \;j \; \frac{E_6}{E_4} = - 2 \pi i E_4^2 E_6
\eta^{-24} \;.\nonumber \eea The logarithmic derivative of
$\eta^{-24}$ is proportional to $E_2$: \beq (\eta^{-24})' = - 2 \pi
i \eta^{-24} E_2 \;. \eeq

As we just saw the derivative of a modular form is not a modular
form, since it does not satisfy eq.~(\ref{modform}) in general. An
exception is the derivative $\partial_T^n F_{1-n}$ which
transforms according to \beq\label{modder}
\partial_T^n F_{1-n}\to (icT+d)^{(n+1)}\partial_T^n F_{1-n}
\eeq and thus is a modular form of weight $n+1$.

\vskip 1cm

%%%%%%%%%%%%%%%%%%%%%%%%%%%%%%%%%%%%%%%%%%%%%%%%%%%%%%%%%%%%%%%%
\noindent {\large\bf Acknowledgments}\\
This work is supported by DFG -- The German Science Foundation -- 
within the ``Schwerpunktprogramm Stringtheorie'' and by
GIF -- the German--Israeli Foundation for Scientific Research, the
European RTN Program HPRN-CT-2000-00148 and the DAAD -- the German
Academic Exchange Service.
We have greatly benefited from conversations with G.~Cardoso and A.~Micu.
\vskip 1cm
%
%%%%%%%%%%%%%%%%%%%%%%%%%%%%%%%%%%%%%%%%%%%%%%%%%%%%%%%%%%%%%%%%
\noindent {\large\bf Note added}\\
After completion of this paper, we became aware of
related work in refs.\ \cite{vH1,vH2}, where similar
phenomena in $N=1$ supersymmetric non-linear
sigma models are analyzed. We thank Jan-Willem van Holten for
drawing our attention to these references.

%%%%%%%%%%%%%%%%%%%%%%%%%%%%%%%%%%%

\end{document}